\documentclass[]{interact}

\usepackage{epstopdf}%
\usepackage[caption=false]{subfig}%

\usepackage[natbibapa,nodoi]{apacite}%
\makeatletter
\NAT@longnamesfalse
\makeatother

\theoremstyle{plain}%

\theoremstyle{definition}

\theoremstyle{remark}

\usepackage{tabularx}
\usepackage{multirow}
\usepackage{makecell}
\usepackage{float}

\usepackage{booktabs}
\usepackage{xcolor}

\usepackage{enumitem}

\newcommand{\pv}[1]{{\scriptsize $#1$}}
\newcommand{\rb}[1]{{\scriptsize ($#1$)}}
\begin{document}

\title{Human Teleoperation for Tele-Ultrasound with Mobile Augmented Reality}
\author{
\name{Elizabeth Klosa\textsuperscript{a}\thanks{CONTACT Elizabeth Klosa. Email: eklosa@student.ubc.ca}, Inzaghi Moniaga\textsuperscript{a}, Victoria Lessoway\textsuperscript{a}, Jessica Li\textsuperscript{b}, David Black\textsuperscript{a}, and Septimiu Salcudean\textsuperscript{a}}
\affil{\textsuperscript{a}Department of Electrical and Computer Engineering, The University of British Columbia, Vancouver, Canada; \textsuperscript{b}Department of Radiology, St. Paul's Hospital, Vancouver, Canada}
}

\maketitle

\begin{abstract}
Access to diagnostic ultrasound remains limited in remote communities where geographical isolation restricts sonographer availability. Human Teleoperation enables sonographers to remotely guide novices wearing mixed-reality headsets to align a physical probe with a virtual probe. However, headset cost limits accessibility, motivating a smartphone-based version. To compensate for missing stereoscopic depth on smartphones, we evaluated visual cues and probe representations. In a study comparing cues, a colour cue most improved positional accuracy, while directional arrows and coordinate axes most improved alignment success. We then compared mobile configurations combining the colour and axes cues against the headset during a simulated scan. Configurations where novices aligned a virtual copy of their probe performed comparably to the headset in alignment error, usability, and image quality, whereas aligning the physical probe reduced usability and image quality. Visual cues and virtual tool representations may therefore make smartphones a viable, lower-cost alternative to headsets for tele-ultrasound.
\end{abstract}


\section{Introduction}\label{intro}
Ultrasound (US) is one of the most widely used medical imaging modalities due to its real-time imaging capability, portability, affordability, and non-invasive, radiation-free nature. However, it is inherently operator dependent, with diagnostic accuracy relying heavily on user expertise~\citep{telerobot_rev}. Combined with a shortage of qualified sonographers, this limits access to high-quality US imaging in rural and remote communities across Canada, which often rely on itinerant sonographers visiting as infrequently as once per month~\citep{Adams2022telerobotic, Davidson2023landscape}, leading to prolonged wait times or long-distance travel for imaging. Access is also unevenly distributed, with low-income and First Nations populations receiving fewer US examinations than the rest of the population~\citep{demographic}. Pregnant women in rural or Indigenous communities are also significantly less likely to receive an obstetric US~\citep{pregnant}. Even where local services exist, access can remain restricted for highly specialized examinations such as fetal echocardiography, offered through only a handful of dedicated programs nationally~\citep{sickkids_program, bcch_cardiology}.

Tele-ultrasound (tele-US) addresses these challenges by enabling experts to remotely guide image acquisition and interpretation in real time~\citep{teleus}. Existing systems fall into two categories. In robotic tele-US, a remote expert directly manipulates the probe through a robotic device, achieving high diagnostic accuracy and reducing reliance on a local operator's skill~\citep{telerobot_rev}. However, it remains limited by cost, complex setup, restricted workspace, on-site support requirements, and human-robot safety challenges~\citep{david_thesis}. In human-guided tele-US, a local operator performs the scan while a remote expert provides real-time guidance, typically through teleconferencing with verbal instruction. This approach achieves diagnostic accuracy comparable to conventional US~\citep{us_comparison}. However, US image quality remains highly operator dependent, with expert guidance shown to significantly improve image quality when scans were performed by inexperienced operators~\citep{sun2022}.

Because precise probe adjustments can be difficult to convey through verbal instruction alone~\citep{teleus_covid_guide, telementored_review}, several mobile tele-US systems have added supplemental visual guidance. Ramsingh et al.~\citeyearpar{ramsingh2019} enabled a remote expert to annotate a smartphone AR video feed during point-of-care US, while Butterfly TeleGuidance similarly overlays visual cues for probe movement~\citep{gillon2024}. However, directing another person's motion through two-dimensional~(2D) annotations or directional cues remains challenging~\citep{black2023human}, since neither explicitly represents the full desired probe position and orientation (pose). Thevenon et al.~\citeyearpar{thevenon2024} addressed this with Echo-QR, representing probe pose using virtual spheres and cubes, though the target pose is fixed and cannot be updated during the exam. None of these approaches provide continuous six degrees of freedom (DoF) AR guidance from a remote expert.

\begin{figure}
    \centering
    \includegraphics[width=1\linewidth]{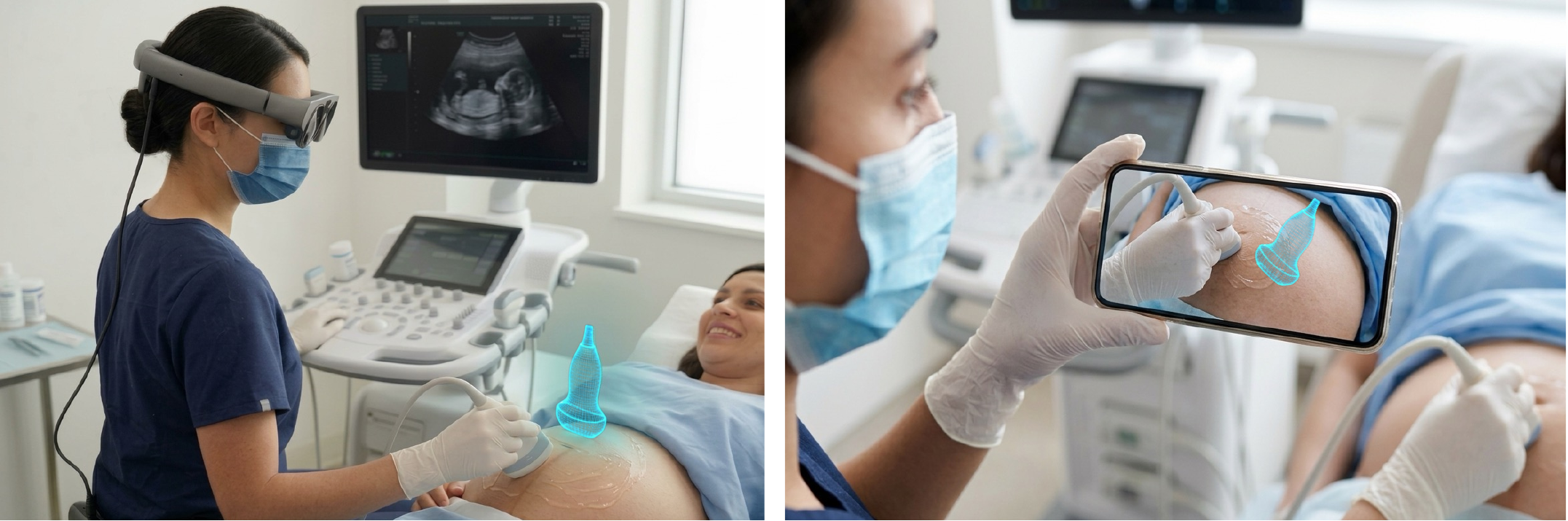}
    \caption{A novice positions an US probe using a mixed reality 3D view (left) and 2D smartphone display (right). We study which visual cues help novices follow a target trajectory. Illustrations were generated with Google Gemini 3.1 Flash Image. The authors confirmed the originality and accuracy of the illustrations, checked the tool's terms of use, and take full responsibility for the content.}    
    \label{fig:ht}
\end{figure}

Human Teleoperation addresses this gap through a tele-US system using mixed reality (MR)~\citep{black2023human}. A local novice follower wears an MR headset and aligns their physical US probe with a virtual probe displayed in the MR environment~(Figure~\ref{fig:ht}). The expert views the live US image stream and specifies the virtual probe pose by manipulating a force feedback haptic input device, allowing them to intuitively guide the follower. This avoids both the cost and workspace constraints of robotic manipulators and the imprecision of verbal or 2D guidance with conventional tele-US. The system has been evaluated in a feasibility study across 754 km~\citep{ryan} and shown to perform comparably to robotic teleoperation with lower setup complexity and cost~\citep{black2025robotic}. While effective~\citep{david_thesis}, MR headsets remain costly for very low-resource settings. Mobile AR is a compelling alternative due to the lower cost and widespread adoption of smartphones~\citep{smartphones,MAR_efficient}. We therefore propose a smartphone-based implementation of Human Teleoperation, in which the virtual probe is displayed on the smartphone screen rather than in a headset, similar to that shown in Figure~\ref{fig:ht}.

Transitioning from a stereoscopic MR headset to a monoscopic handheld device changes virtual content perception. Headsets provide stereoscopic rendering and head-coupled motion parallax, whereas smartphones render from the device camera rather than the user's viewpoint, so users must rely primarily on monocular depth cues such as occlusion, relative size, and shading~\citep{cooper_percept,kruijff}. Consequently, handheld distance judgments can be inconsistent, with virtual objects perceived as closer in one study~\citep{dey_tablet} yet farther away in another~\citep{liu_handheld}.

With a handheld device, attention is divided between the screen and surroundings because virtual content is not integrated into the user's environment. This has been shown to degrade performance in mobile AR navigation relative to both an MR headset and an unaided control group~\citep{navaids}. However, the impact of attention division may depend on task scale. In a close-range CT-guided needle placement task, Li et al.~\citeyearpar{li_needle} found comparable target overlay error between a smartphone and an MR headset. The smartphone required longer placement times but produced slightly lower final placement error, which the authors attributed to the headset's vergence-accommodation conflict, an effect that is more pronounced at close range. This suggests attention division may matter less at close range, where a monoscopic display avoids the perceptual issues associated with stereoscopic displays~\citep{kramida2016vac}.

Visual cues can further support spatial alignment, though prior work using monitors and headsets has produced heterogeneous findings. Object-centric arrow-based cues outperformed triplanar and sight-alignment displays for 5-DoF and 6-DoF alignment on a 2D monitor~\citep{trac}. A coordinate-axes cue was fastest and most accurate for alignment using an MR headset, while a reticle cue improved rotational accuracy at the cost of speed and translational accuracy~\citep{andersen2020ar}. A directional arrow performed well, with numerical or colour-deviation cues providing no additional benefit despite their addition being preferred by participants~\citep{deviation}. A colour-coded residual-error cue matched a no-feedback condition, but outperformed an exaggerated shape cue~\citep{fidalgo2025}.  A goal-based cue, conveying the target pose through shape, also outperformed an action-based cue showing the required movement, with their combination performing worse than either alone despite being preferred by participants~\citep{liu6dof}. No cue design was consistently superior across studies.

Prior work on handheld AR has considered cue design and pose guidance, but these have generally been studied separately. Wacker et al.~\citeyearpar{wacker} compared four position-visualization methods for a mid-air pen on a smartphone and found little difference in accuracy or speed, though a heatmap was most preferred. Wieland et al.~\citeyearpar{wieland} found a three-dimensional~(3D) arrow to be fastest and best rated for guiding users to out-of-view objects, though in a large-scale search task rather than close-range guidance. Handheld AR has been used for single-target pose guidance, such as needle insertion~\citep{hecht2020,needleguide2025}, though toward a fixed target rather than the continuous, dynamic alignment required for an US scan. To our knowledge, no prior study has evaluated visual guidance cues for close-range, dynamic 6-DoF AR alignment on a handheld display, leaving open which cue designs best support this task and whether findings from monitors and headsets generalize to this setting.

In this work, we present a smartphone-based implementation of Human Teleoperation for US guidance. Its design is motivated by the known limitation of handheld AR, where a phone screen cannot show stereoscopic depth, forcing users to rely on monocular cues. We therefore expected depth to be the main source of alignment error, and designed each guidance cue to help users judge depth and probe pose.

In the original system, the follower aligns their physical probe with the expert's virtual probe, which we refer to as real-to-virtual~(R-V) alignment. On a smartphone, this becomes more difficult since the follower must judge the depth of a virtual object relative to a real one on a monoscopic display. This led us to also consider virtual-to-virtual~(V-V) alignment, in which the follower aligns a tracked virtual copy of their own probe with the expert's virtual probe. 

We evaluate the system in two user studies. The first compares four guidance cues (Colours, Arrows, Axes, and Projection) against a baseline condition to identify which best supports R-V alignment during dynamic probe tracking (Section~\ref{study1_guidancecues}). The second study incorporates the best-performing cues from the first study into three smartphone-based configurations that vary the follower probe representation and background (Section~\ref{study2_implementation}), namely Mobile R-V, Mobile V-V with the camera feed displayed (Feed), and Mobile V-V without it (No Feed). These are compared against the original Magic Leap (ML) system in an obstetric phantom scanning study to determine whether the smartphone-based configurations can perform similarly to the headset, and to examine the effect of follower probe representation.

\section{Methods}\label{sec2}
Both studies were approved by the University of British Columbia Behavioural Research Ethics Board, and all participants provided informed consent prior to participation.

\subsection{Human Teleoperation System}\label{HT_overview}

\begin{figure}[h!]
    \centering
    \includegraphics[width=1\linewidth]{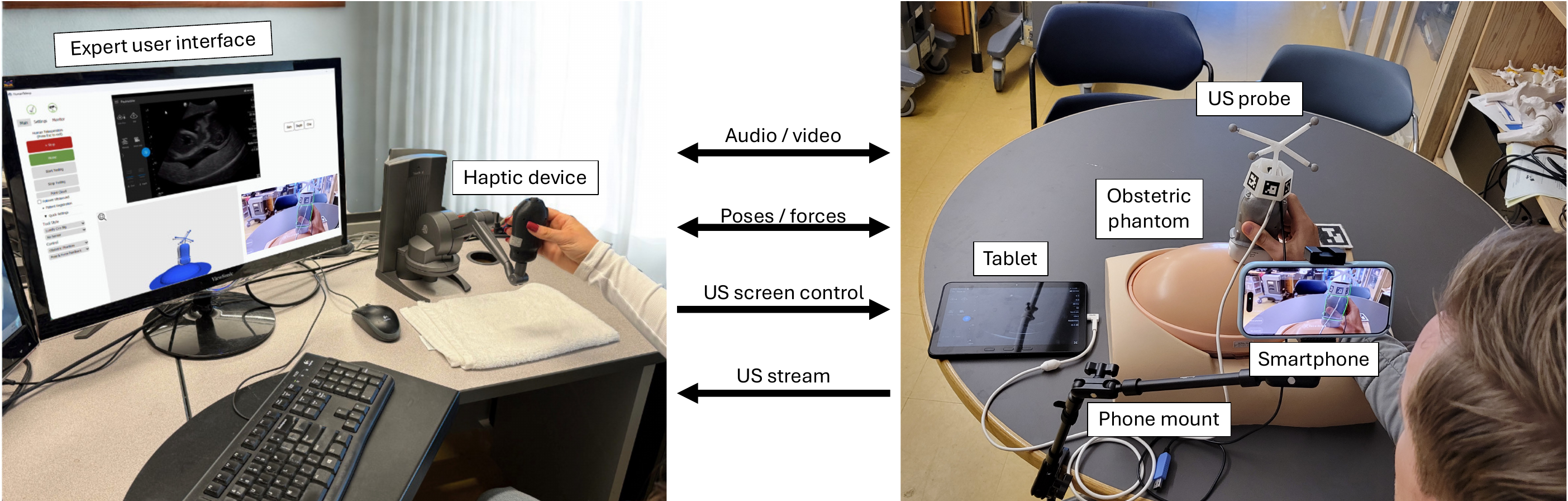}
    \caption{Overview of Human Teleoperation and information exchange. The expert (left) uses a haptic device and views a user interface. The follower (right) uses a smartphone in a mount, an US probe, a tablet, and an obstetric phantom.}
    \label{fig:human_teleop_sides}
\end{figure}
Details of the MR headset-based Human Teleoperation system are provided in previous work~\citep{black2023human,teleop_comm,david_thesis}. The mobile system retained the same expert-side setup and communication architecture, with the main change being the follower's headset replaced by a smartphone to display the virtual US probes and guidance cues. The follower mobile application was developed in Unity 2022.3.62f1 using the AR Foundation framework\footnote{https://docs.unity3d.com/Packages/com.unity.xr.arfoundation@5.2/manual/index.html}, which provides access to the Apple ARKit and Google ARCore XR Plug-ins, and was implemented for both iOS and Android. Study 1 used a Samsung Galaxy S21+ 5G smartphone and Study 2 used an iPhone 16 Pro. The follower used a Philips Lumify Curved Array Transducer (C5-2) in both studies. 

To position guidance relative to the real probe and measure how closely the follower matched the expert, the follower probe pose had to be tracked. The probe was fitted with a 3D-printed hexagonal fixture (Figure~\ref{fig:human_teleop_sides}) that supported two tracking methods. Six 2 cm $\times$ 2 cm ArUco markers~\citep{garrido2014aruco} on its faces allowed camera-based ArUco tracking from the smartphone, and a configuration of reflective infrared markers on the top allowed for optical tracking with an external tracker. 

For ArUco tracking, a custom OpenCV plugin in Unity detected DICT\_5X5\_50 markers in the device RGB frames. The six markers on the fixture defined an ArUco marker board with known relative geometry, and a single 6-DoF pose of the fixture relative to the phone camera was estimated using perspective-$n$-point (PnP). Both studies used this fixture and detection pipeline, but they differed in how virtual content was anchored and in which tracking method provided the probe pose used for guidance and evaluation.

In Study 1, AR tracking on the Samsung Galaxy S21+ was prone to world-frame drift, so virtual content was anchored to the camera and the phone was held stationary in a fixed holder. ArUco tracking was not yet stable enough to drive the guidance, so optical tracking provided the probe pose. The reflective infrared markers were tracked by a Northern Digital Inc. (NDI) Polaris Spectra optical tracker positioned opposite the follower, providing the ground-truth pose. A rigid transform from the NDI frame into the camera frame was computed at the start of each session using simultaneous ArUco and NDI measurements, after which NDI poses were mapped into camera coordinates and applied to the virtual follower probe. The NDI pose measurements were thus used to generate visual feedback and quantify the follower's pose error. See further discussion in Section~\ref{study1_limitations}.

In Study 2, the iPhone 16 Pro did not exhibit this drift, so virtual content was anchored in the world frame and the phone could move freely during scanning. ArUco tracking had become stable enough on its own at this stage, so the ArUco estimate was used directly to compute the reported pose errors and position the guidance cues. Each pose was transformed into the Unity world frame, and display poses were temporally smoothed using linear and spherical interpolation. A separate ArUco marker attached to the obstetric phantom was detected in the same frames to estimate the phantom pose, which determined the placement of the virtual phantom in the Unity AR scene and its registration in the expert's haptic workspace.

The virtual probe was rendered as a semi-transparent silver model ($\alpha=0.5$), allowing the physical environment to remain visible through it. To improve probe shape perception, silhouette outlines were rendered along its edges (Figure~\ref{fig:study1_visuals}) using a custom edge-detection shader that detects depth and normal discontinuities with the Roberts cross operator, adapted from an existing Unity technique\footnote{https://ameye.dev/notes/edge-detection-outlines/}. The specific cue renderings are described in Section~\ref{study1_guidancecues} and Section~\ref{study2_implementation}.

The follower probe was connected to a tablet running a custom application through which the US feed was streamed to the expert, who also had remote control of the tablet interface to adjust imaging parameters and annotate the US images~\citep{black2024userinteraction}. The expert controlled the pose of the virtual probe through a Touch X haptic device (3D Systems, Rock Hill, SC), a desktop serial manipulator that provides 6-DoF input and 3-DoF force feedback~(Figure~\ref{fig:human_teleop_sides}). The device was fitted with a 3D-printed probe handle to mimic the sensation of a real US scan. Guided by the follower's real-time US feed, the expert continually adjusted the handle pose, which updated the virtual probe pose on the follower's screen. The expert could additionally view the follower's screen along with both probes and the phantom rendered in the haptic workspace. Communication between the follower and expert sides was achieved using WebRTC, a peer-to-peer framework supporting low-latency communication across network types (Ethernet, Wi-Fi, 4G LTE, and 5G)~\citep{teleop_comm, black2023human}.

\subsection{Study 1: Guidance Cue Comparison}
This study compared four guidance cues and a baseline condition for supporting probe alignment within our mobile Human Teleoperation system through a user study consisting of a series of alignment tasks. Guidance cue design was evaluated independent of live expert input, so pre-recorded expert probe trajectories were streamed to the follower device through the WebRTC pipeline rather than controlled live.

\subsubsection{Experimental Setup}
The follower setup is shown in Figure~\ref{fig:human_teleop_sides}, although a compliant material replaced the obstetric phantom as a scanning surface, since participants followed a virtual US probe target rather than performing an actual US scan. The follower sat at a desk equipped with a fixed phone holder to ensure consistent camera positioning across trials. The workspace surface was pre-scanned before each trial to ensure reliable AR plane detection for Projection.

\subsubsection{Guidance Cues}\label{study1_guidancecues}
\begin{figure}[!h]
    \centering
    \includegraphics[width=1\linewidth]{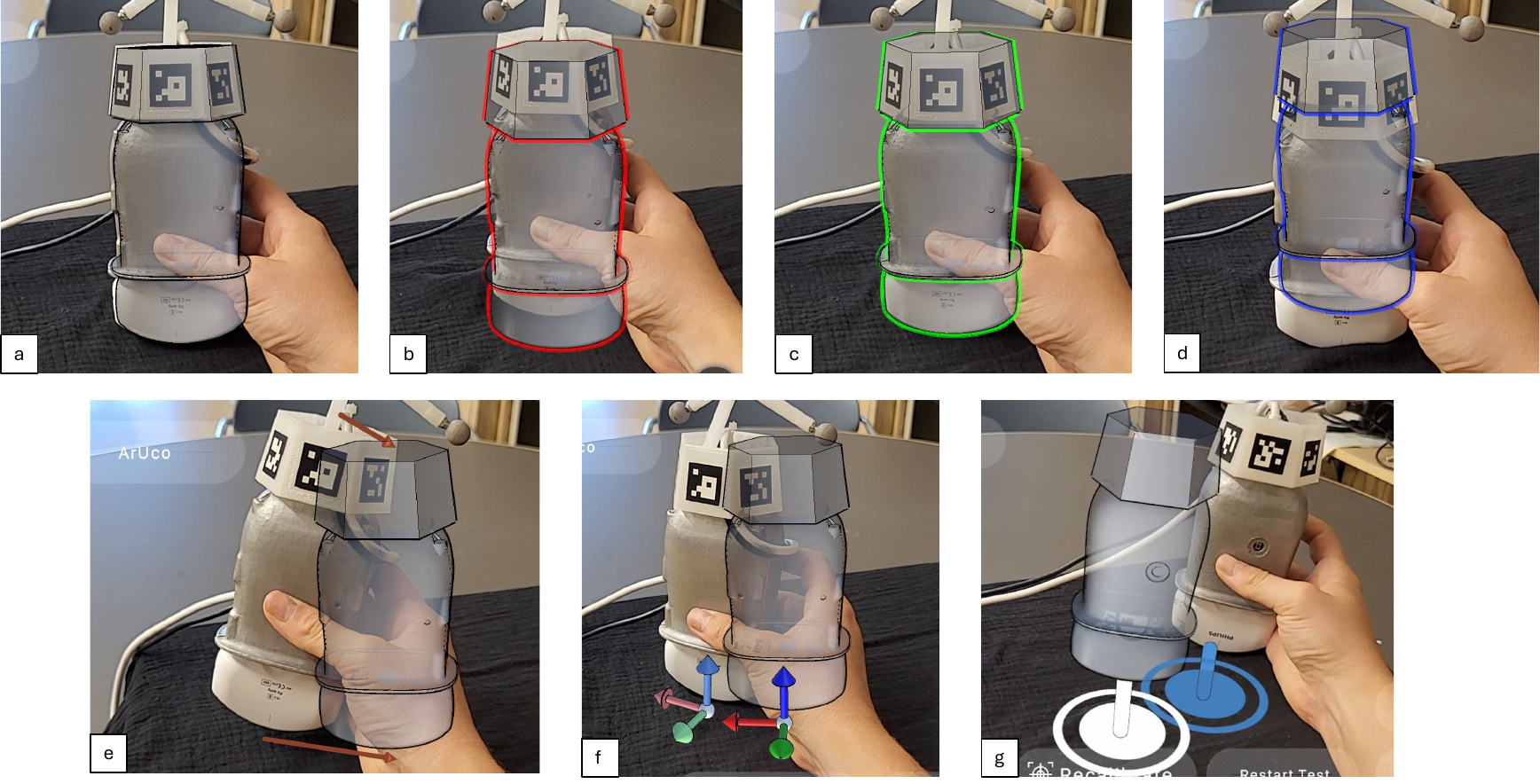}
    \caption{Baseline and guidance cue conditions. a) Baseline. b) Colours (follower too far). c) Colours (aligned). d) Colours (follower too close). e) Arrows. f) Axes. g) Projection.}
    \label{fig:study1_visuals}
\end{figure}

The baseline condition showed only the virtual expert probe, while each of the other four conditions added one guidance cue to it. These conditions are described below and shown in Figure~\ref{fig:study1_visuals}.

\textbf{Baseline:} The virtual expert probe was shown without any added cue~(Figure~\ref{fig:study1_visuals},~a).

\textbf{Colours:} An outline of the virtual expert probe changed colour based on the alignment error between the expert and follower probes in the depth axis, defined as the camera's forward axis. Within 5~mm, the probes were considered aligned and the outline remained green (Figure~\ref{fig:study1_visuals},~c). As the follower moved too close or too far behind the expert, the colour faded to blue or red, respectively, indicating a small misalignment. Beyond 10~mm, the expert outline appeared fully blue (Figure~\ref{fig:study1_visuals},~d) or red (Figure~\ref{fig:study1_visuals},~b). The 5~mm and 10~mm thresholds were chosen empirically to give distinguishable feedback while avoiding excessive colour changes from small movements.

\textbf{Arrows:} Arrows at both the top and the bottom tips of the probe (Figure~\ref{fig:study1_visuals}, e) indicated the direction of movement required to align both ends of the follower probe with the expert probe. Each arrow pointed along the 3D vector from the follower probe tip to the corresponding expert probe tip, with its length scaling with the distance between them, and was rendered as a 3D object in the scene. Each arrow disappeared once the corresponding follower tip was within 5~mm.

\textbf{Axes:} A 3D coordinate axes (red, green, blue for X, Y, Z) was attached to the front of each probe~(Figure~\ref{fig:study1_visuals},~f), with the follower probe's axes rendered at reduced colour intensity to distinguish it from the expert probe. This allowed users to compare the orientation of the two probes directly.

\textbf{Projection:} An AR Raycast Manager projected a vertical ray downward from both the follower and expert probe tips. When the rays intersected a detected AR plane, reticles were placed and a line was rendered to connect them to the probe tips~(Figure~\ref{fig:study1_visuals},~g). This provided users with a spatial reference relative to the environment to aid depth perception.

\subsubsection{User Study Design}\label{study1_userstudy}
Twelve participants ($n=12$; 5 female, 7 male; aged 24--40) with varying levels of AR experience (rated on a 5-point scale, mean = 3.1 $\pm$ 1.5) took part in the study, each completing one trial per condition, preceded by an unrecorded practice trial to familiarize them with the system and tasks. In each trial, participants followed a prerecorded expert trajectory displayed on the mobile device by aligning their physical probe with the virtual expert probe. 

Each trial consisted of five move-and-hold segments, each comprising a dynamic and a steady-state phase, evaluating dynamic tracking during movement and alignment accuracy while stationary. In each segment, the expert probe moved to a target pose and then remained stationary for approximately 10--15 seconds before proceeding to the next target. Participants completed six trials total, with one practice trial and one trial for the baseline and each guidance cue. The same set of trajectories was used across participants, with trajectory assignments and condition order randomized to minimize potential learning effects and bias due to differences in trajectory difficulty. Each trial lasted approximately two minutes.

\subsubsection{Data Analysis}\label{study1_DA}
The effectiveness of each cue was evaluated using both quantitative and qualitative measures. Translational error was quantified as the mean absolute position error along each axis (X, Y, and Z) and the overall positional error, calculated as the mean Euclidean distance between the origins of coordinate frames rigidly attached to the expert and follower probes, located at the bottom tip of each probe. Per-axis positional error was expressed in a device-aligned frame, where X is left-right on the screen, Y is up-down, and Z is the viewing axis out of the phone toward the workspace. The Z-axis was of particular interest due to the known difficulty of depth perception in monocular handheld AR. Rotational error was quantified as the geodesic angular error, calculated as the magnitude of the relative rotation between the expert and follower probe frames. 

All error metrics were computed separately for dynamic and steady-state phases. Steady-state alignment was further characterized by success rate, where a hold was considered successful if the follower probe remained within 5~mm and 5$^\circ$ of the expert pose for at least three continuous seconds during the steady-state phase. These thresholds were chosen to require alignment close enough to reproduce the expert's intended view while remaining achievable for novice followers, and are comparable to the accuracy previously achieved by followers using the MR headset-based system~\citep{black2023humanasrobot}. The three-second hold ensured that success reflected a stable view rather than the probe briefly passing through the window. Success rate was the proportion of successful holds (out of five per trial) per participant and condition. 

Settling time was defined as the time from the onset of the steady-state phase until the follower probe first remained within the success window for three continuous seconds, and was therefore defined only for successful holds. Each participant's settling time for a given condition was the mean across their successful holds, providing a measure of how quickly participants converged. 

Participants also completed a survey on the usability, perceived usefulness, and overall preference of each condition. Results are reported as mean $\pm$ standard deviation (SD) across participants. Non-parametric tests were used given the small sample size. Each cue was compared to the Baseline using Wilcoxon signed-rank tests, with the matched-pairs rank-biserial correlation reported as the effect size ($r_{\mathrm{rb}}$), interpreted using Cohen's conventional thresholds (approximately 0.10, 0.30, and 0.50 for small, moderate, and large, respectively)~\citep{cohen1988}. A positive $r_{\mathrm{rb}}$ indicates higher values than Baseline and a negative $r_{\mathrm{rb}}$ indicates lower values, so negative values reflect improvement for error and settling time, while positive values reflect improvement for success rate. Holm correction was applied across the four comparisons within each outcome measure and phase, rather than across all outcomes combined, with significance set at $\alpha=0.05$.

\subsection{Study 2: Mobile vs.\ Magic Leap Comparison}\label{ML_vs_mobile}
To compare the proposed mobile approach with the original MR headset-based system, a comparative study was conducted using the Magic Leap and three mobile configurations. In one, participants aligned the physical probe with the virtual expert probe. In the other two, they aligned a virtual copy of their probe, either overlaid on the live camera feed or on a black background. The study was designed to assess whether a mobile approach could achieve performance similar to the original MR system and to examine how different visual representations affected performance.

\begin{figure}
    \centering
    \includegraphics[width=1.0\linewidth]{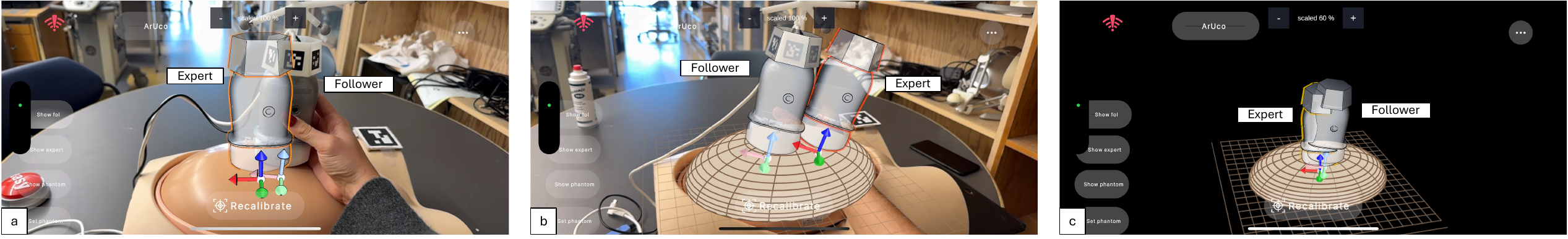}
    \caption{a) Mobile R-V. b) Mobile V-V Feed. c) Mobile V-V No Feed.}
    \label{fig:v_v}
\end{figure}

\subsubsection{System Implementation}\label{study2_implementation}
The MR system used in this study is introduced in Section~\ref{intro} and described further by~\citet{david_thesis}. The Magic Leap 2 was used, with the headset's built-in ArUco tracking providing probe pose tracking. No visual feedback cues were added to the Magic Leap's virtual scene.

Based on the findings of Study 1, the Colours and Axes cues were combined in all three mobile configurations to provide complementary depth and pose guidance. Configurations included Mobile R-V, Mobile V-V Feed, and Mobile V-V No Feed. Here, R-V denotes alignment of the real physical follower probe with the virtual expert probe, while V-V denotes alignment of the virtual follower probe, rendered directly on top of the real probe according to its tracked pose, with the virtual expert. 

Feed refers to a live camera feed that was retained as the background, with the virtual follower probe, expert probe, and obstetric phantom overlaid on their corresponding real-world objects~(Figure~\ref{fig:v_v}, b). It should be noted that the R-V configuration also retained the live camera feed~(Figure~\ref{fig:v_v}, a), but does not render the virtual follower probe or virtual phantom. No Feed used the same virtual representations but removed the camera feed and reduced the virtual objects to 60\% of their original size (Figure~\ref{fig:v_v}, c) to allow the phone to move closer to the workspace while still viewing the whole virtual scene. The background was rendered black, replacing the camera feed to reduce visual clutter. The follower could still see the phantom and their US probe around the phone. 

For both V-V configurations, the virtual obstetric phantom was rendered with concentric circles to provide an additional cue to the 3D surface curvature. As the virtual follower and expert probes approached and contacted the phantom, they were occluded by its surface, and the circle pattern was similarly occluded by the probes. This allowed users to judge the probe position relative to the phantom surface based on where the circle lines were hidden. A surface grid was also rendered beneath the phantom as a supplementary perspective cue.

The Colours cue was modified for this study to encode the magnitude of the depth error rather than its direction. The outline remained green when the follower probe was within 5~mm of the expert probe along the depth axis, faded to orange as the error grew beyond 5~mm, and appeared fully red beyond 10~mm in either direction. In Study 1, participants reported difficulty recalling whether blue or red meant too close or too far, whereas the green-to-orange-to-red progression follows a more familiar and intuitive ordering, while still providing meaningful depth feedback.

\subsubsection{Experimental Setup}\label{study2Setup}
To facilitate the experimental setup, the follower and expert were situated in the same room, but were prevented from seeing or communicating with each other. A novice follower was positioned in front of a Space Fan-ST obstetric phantom~(US-7A, Kyoto Kagaku) simulating a second-trimester pregnant abdomen with an internal fetal model to support realistic US scanning tasks (Figure~\ref{fig:human_teleop_sides}). The fetal model was maintained in a fixed head-down, spine-down position throughout the study to ensure consistency across participants and scans.

For Mobile R-V, participants held the smartphone in their non-dominant hand and the probe in their dominant hand. This follows the setup in Figure~\ref{fig:human_teleop_sides}, with the exception that the phone mount was removed and the device was held directly by the participant, allowing them to reposition it as needed to keep the relevant workspace in view. However, this limited participants' ability to adjust their grip on the probe and interact with the device. Because the ability to reposition the phone's view was rarely used, the smartphone was instead secured in a holder for the two V-V configurations, freeing both hands for participants to reposition the probe, adjust their grip, and interact with the device as needed.

\subsubsection{Obstetric Haptic Model}
Haptic feedback was generated using a geometric approximation of the obstetric phantom, consisting of the upper half of an ellipsoid resting on a bounded planar brim. The coordinate frame was centered at the ellipsoid center, $\mathbf{x}_{c}$, with the brim plane at $y=0$ and the dome occupying $y\geq0$. The expert probe position in this frame is given by $\mathbf{x}_{e}=R^{\top}(\mathbf{x}_{w}-\mathbf{x}_{c})=[x,\,y,\,z]^{\top}$, where $\mathbf{x}_{w}$ is the world frame probe position and $R$ is the world frame phantom orientation. The full ellipsoid is defined as

\begin{equation}
\Phi(\mathbf{x}) = \frac{x^{2}}{a^{2}} + \frac{y^{2}}{b^{2}}
+ \frac{z^{2}}{c^{2}} = 1,
\end{equation}
with semi-axes set as $a = 146$~mm, $b = 60$~mm and $c = 107.5$~mm based on measurements of the physical phantom.

Penetration was calculated separately for the dome and brim regions and blended across a transition band of 6~mm about the brim plane using a cubic smooth-step function so that the two surfaces remained continuous, giving the blended penetration vector $\boldsymbol{\delta}$. Dome contact was detected when $\Phi(\mathbf{x}_{e})<1$, with $y$ clamped at the brim plane, following the ellipsoid penetration model described by~\citet{ryan}. The contact direction was taken as the normalized ellipsoid gradient at $\mathbf{x}_{e}$, and the penetration depth was determined from the smallest positive root of the intersection of a ray along this direction with the ellipsoid surface.

The brim was modelled as a bounded plane. The brim term included a displacement along $+\hat{\mathbf{y}}$ equal to the depth of the probe below the brim plane. Within the dome footprint, the brim term additionally supplied a radial displacement toward the rim. 

The resulting penetration vector $\boldsymbol{\delta}$ was rendered using a spring-damper model,
\begin{equation}
\mathbf{f} = R\left[k_{p}\boldsymbol{\delta} - b_{p}\dot{x}_{e,t}\right],
\qquad k_{p} = 0.5~\mathrm{N\,mm^{-1}},
\end{equation}
and rotated into the haptic workspace, where $\dot{x}_{e,t}$ is the probe velocity tangential to the contact normal and the damping constant $b_{p} = 0.0015$~N\,s\,mm$^{-1}$. No additional damping was applied along the contact normal on the dome to avoid a touchdown force step and adhesion on release. An additional damper of the same coefficient acted along $\hat{y}$ at or below the brim plane, blended across the transition band, coinciding with the contact normal on the flat brim but not on the dome. The rendered force magnitude was limited to $7.5$~N.

\subsubsection{User study}
Sixteen participants in total took part as followers across two days, with twelve on each day and eight completing both days (Day 1: $n=12$, 8 female, 4 male; aged 20--69. Day 2: $n=12$, 8 female, 4 male; aged 20--29). On the first day, each participant completed the guided obstetric US scan using the Mobile R-V configuration and the Magic Leap system. On the second day, participants completed the same scan using the two Mobile V-V configurations. Within each day, the order of the two conditions was alternated across participants to reduce potential ordering effects. All scans were guided by a single expert sonographer with over 30 years of US experience.

\subsubsection{Procedure}\label{study2_procedure}
Each scan represented a shortened version of a clinical obstetric US examination, with the scope determined by the participating expert sonographer to keep the study duration manageable. During each scan, the expert sonographer guided participants to locate and acquire the following anatomical structures and measurements:

\begin{enumerate}[label=Task \arabic*:, leftmargin=4em, align=left]
\item Document fetal presentation (cephalic or breech)
\item Document the sagittal placental location (lower tip)
\item Document the transverse placental location
\item Measure the deepest vertical pocket of amniotic fluid
\item Document fetal stomach
\item Document fetal bladder
\end{enumerate}

\subsubsection{Data Analysis}\label{study2_dataanalysis}
The three mobile configurations were evaluated against the Magic Leap using quantitative and qualitative measures. Translational error was calculated as the mean Euclidean distance between the origins of coordinate frames rigidly attached to the expert and follower probes, located at the bottom tip of each probe. Rotational error was defined as the geodesic angular magnitude of the relative rotation between the expert and follower probe frames. 

Per-task centered errors were also calculated by removing the mean follower-expert position offset and mean relative rotation within each task before computing the residual position norms and rotation angles. Per-participant centered errors were then calculated as the sample-weighted mean of these residuals across tasks. 

Active scanning duration was defined as the total time within each task, excluding periods when the stream was frozen while the expert analyzed the image, since these periods did not depend on the follower's probe alignment performance. Overall duration was computed as the sum of the active scanning durations across the tasks.

Each mobile configuration was compared with the Magic Leap using the same statistical analysis as the previous study (Section~\ref{study1_DA}). Here, $r_{\mathrm{rb}}$ was signed so that a positive value indicates worse performance for the mobile configuration than for the Magic Leap, corresponding to higher pose error or duration, or lower image quality. Mobile R-V and Magic Leap were tested in the same session ($n=12$). The two V-V configurations were tested in a separate session, so comparisons with Magic Leap were limited to the $n = 8$ participants who completed both sessions. Holm correction was applied separately to the overall and task-level comparisons for each metric. Overall comparisons were corrected across the three mobile configurations, and task-level comparisons were corrected across the six tasks within each mobile configuration. Pose error and duration were treated as separate correction families ($\alpha=0.05$). After each session, participants completed the Raw NASA Task Load Index (TLX) and System Usability Scale (SUS), and rated the quality of the AR guidance on a 5-point Likert scale. Participants who completed both sessions also ranked all four conditions by preference.

\begin{table}[H]
\centering\scriptsize
\setlength{\tabcolsep}{3pt}
\caption{Image quality rating scale.}
\label{tab:quality_scale}
\begin{tabularx}{\columnwidth}{Xc}
\toprule
Score Definition & Score \\
\midrule
Not obtained & 0 \\
Image quality is too poor to permit meaningful interpretation & 1 \\
Image quality is better than poor but less than suboptimal for interpretation & 2 \\
Suboptimal image quality, but basic image interpretation is possible & 3 \\
Better than suboptimal but not good & 4 \\
Good image quality, meaningful image interpretation is easy & 5 \\
\bottomrule
\end{tabularx}
\end{table}

Since the ultimate goal is to support real-world US scans, image quality was assessed alongside tracking accuracy to evaluate actual scanning performance. Images were rated by a radiologist using a six-point rating scale (Table~\ref{tab:quality_scale}), which was previously introduced and validated~(\citeauthor{millington2016race}, \citeyear{millington2016race}; \citeauthor{millington2017acts}, \citeyear{millington2017acts}), demonstrating good inter-rater reliability and the ability to detect changes in learner performance. Images were anonymized and randomized, with approximately 9\% of images repeated within the rating set to assess intra-rater reliability. The radiologist was blinded to condition. Reliability was quantified using exact agreement, agreement within one point, and quadratic-weighted Cohen's kappa. Scores were averaged per participant and condition. Each mobile configuration was compared with Magic Leap using the same Wilcoxon signed-rank and Holm correction procedure used for the error and duration analyses.

\section{Results}\label{results}
\subsection{Study 1: Guidance Cue Comparison}

\begin{figure}[H]
    \centering
    \includegraphics[width=1.0\linewidth]{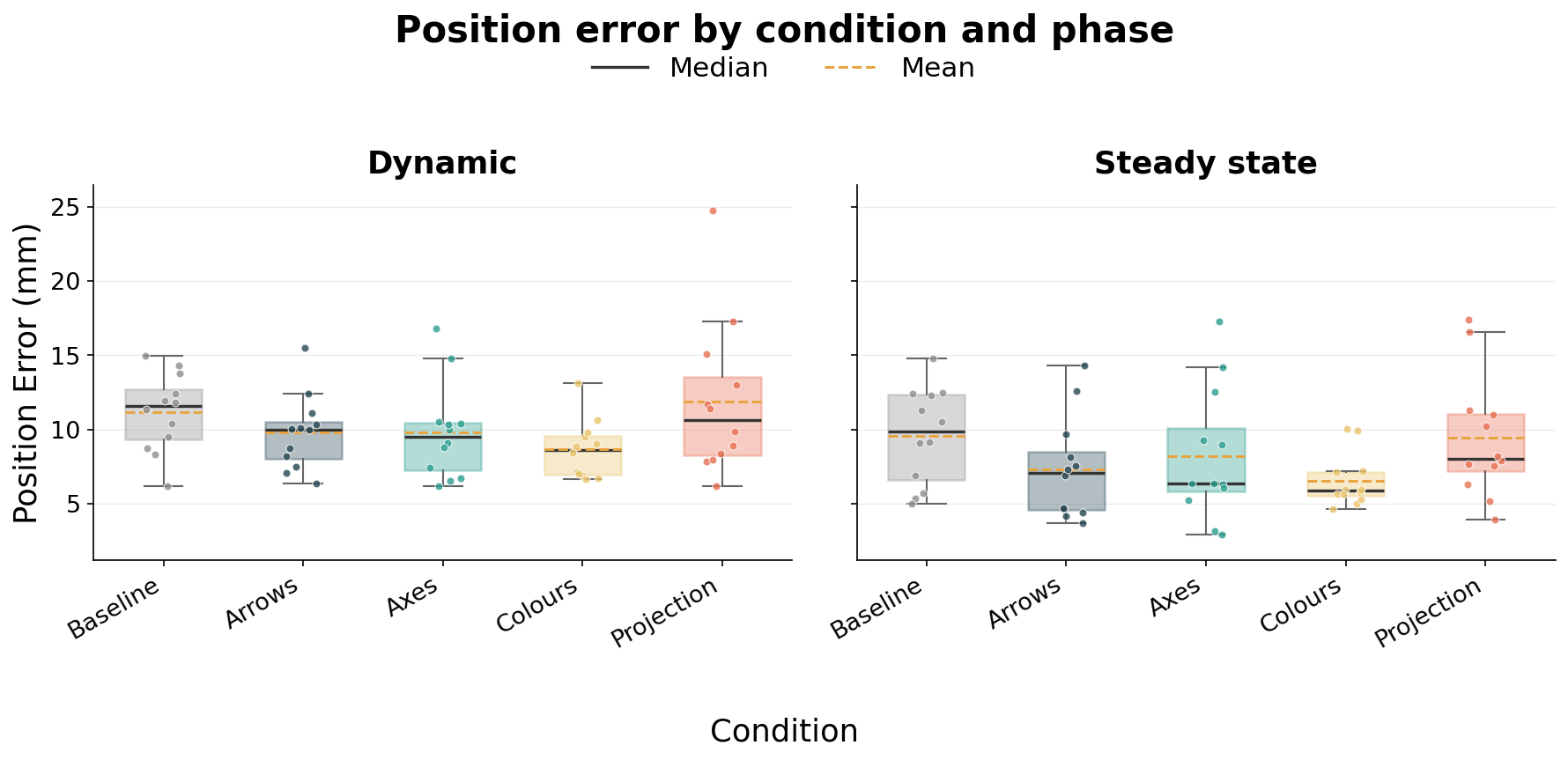}
    \caption{Positional error by condition for the dynamic and steady-state phases. Each point represents one participant's mean error. Boxes show the interquartile range.}
    \label{fig:pos}
\end{figure}

We evaluated four guidance cues and a baseline condition across twelve participants. The mean positional error by condition and phase is shown in Table~\ref{tab:study1_pos_errors} and Figure~\ref{fig:pos}. Colours produced the lowest mean positional error of any condition in both phases and was the only cue to improve significantly upon the Baseline, reducing mean positional error by 2.5~mm in the dynamic phase ($p = 0.027, r_{\mathrm{rb}} = -0.85$) and by 3.1~mm in the steady-state phase ($p = 0.037,  r_{\mathrm{rb}} = -0.82$). Arrows and Axes also reduced positional error relative to Baseline, although neither reached significance. Despite this, Arrows showed a large effect in both phases ($r_{\mathrm{rb}} = -0.64$), while Axes showed a large effect in the dynamic phase ($r_{\mathrm{rb}} = -0.64$) and moderate effect in the steady-state phase ($r_{\mathrm{rb}} = -0.38$). Projection was comparable to Baseline, with small effect sizes in the dynamic and steady-state phases ($r_{\mathrm{rb}} = +0.21$ and $-0.08$, respectively). 

Across every condition, the Z-axis carried the highest error. Colours significantly reduced Z-axis error in both the dynamic and steady-state phases ($p = 0.010$, $r_{\mathrm{rb}} = -0.92$ and $p = 0.006$, $r_{\mathrm{rb}} = -0.95$, respectively). Z-axis error was significantly reduced by Axes in the dynamic phase ($p = 0.010$, $r_{\mathrm{rb}} = -0.92$) and Arrows in the steady-state phase ($p = 0.028$, $r_{\mathrm{rb}} = -0.82$), with large effect sizes.

\begin{table}[h!]
\centering\scriptsize
\setlength{\tabcolsep}{3pt}
\caption{Mean per-axis and overall positional errors ($\pm$ SD) for the five conditions, by phase. Holm-corrected paired Wilcoxon $p$-values and rank-biserial correlations $r_{\mathrm{rb}}$ are given for each cue against Baseline, for the overall position error and for the Z-axis error alone. Positive $r_{\mathrm{rb}}$ indicates higher error than Baseline. Significant values ($p<0.05$) are bold.}
\label{tab:study1_pos_errors}
\resizebox{\textwidth}{!}{%
\begin{tabular}{@{}cl|ccc|c|cc|cc@{}}
\toprule
\textbf{Condition}& \textbf{Phase} & \multicolumn{3}{c|}{\textbf{Position (mm)}} 
& \makecell{\textbf{Avg. pos} \\[-3pt] \textbf{error (mm)}} 
& \multicolumn{2}{c|}{\textbf{Overall vs Baseline}} 
& \multicolumn{2}{c}{\textbf{Z only vs Baseline}} \\[-2pt]
 & & X & Y & Z &  & $p$ & $r_{\mathrm{rb}}$ & $p$ & $r_{\mathrm{rb}}$ \\
\midrule
\multirow{2}{*}{\textbf{Baseline}} & Dynamic & 4.6& 3.7& 8.2& 11.2 $\pm$ 2.6& -- & -- & -- & -- \\
 & Steady-state & 2.6& 3.8& 7.8& 9.6 $\pm$ 3.3& -- & -- & -- & -- \\
\midrule
\multirow{2}{*}{\textbf{Arrows}} & Dynamic & 4.4& 3.8& 6.5& 9.8 $\pm$ 2.5& 0.16 & $-0.64$ & 0.054 & $-0.72$ \\
 & Steady-state& 2.7& 3.6& 4.9& 7.3 $\pm$ 3.4& 0.16 & $-0.64$ & \textbf{0.028} & $-0.82$ \\
\midrule
\multirow{2}{*}{\textbf{Axes}} & Dynamic & 4.8& 3.5& 6.1& 9.8 $\pm$ 3.2& 0.16 & $-0.64$ & \textbf{0.010} & $-0.92$ \\
 & Steady-state& 2.4& 3.1& 6.5& 8.2 $\pm$ 4.4& 0.53 & $-0.38$ & 0.53 & $-0.38$ \\
\midrule
\multirow{2}{*}{\textbf{Colours}} & Dynamic & 4.5& 3.0& 5.2& 8.7 $\pm$ 2.0& \textbf{0.027} & $-0.85$ & \textbf{0.010} & $-0.92$ \\
 & Steady-state& 3.0& 3.0& 3.6& 6.5 $\pm$ 1.8& \textbf{0.037} & $-0.82$ & \textbf{0.006} & $-0.95$ \\
\midrule
\multirow{2}{*}{\textbf{Projection}}& Dynamic & 4.9& 4.9& 7.7& 11.9 $\pm$ 5.2& 0.57 & $+0.21$ & 0.30 & $-0.36$ \\
 & Steady-state& 2.3& 4.4& 6.7& 9.4 $\pm$ 4.1& 0.85 & $-0.08$ & 0.53 & $-0.28$ \\
\bottomrule
\end{tabular}%
}
\end{table}
\begin{figure}[h!]
    \centering
    \includegraphics[width=0.95\linewidth]{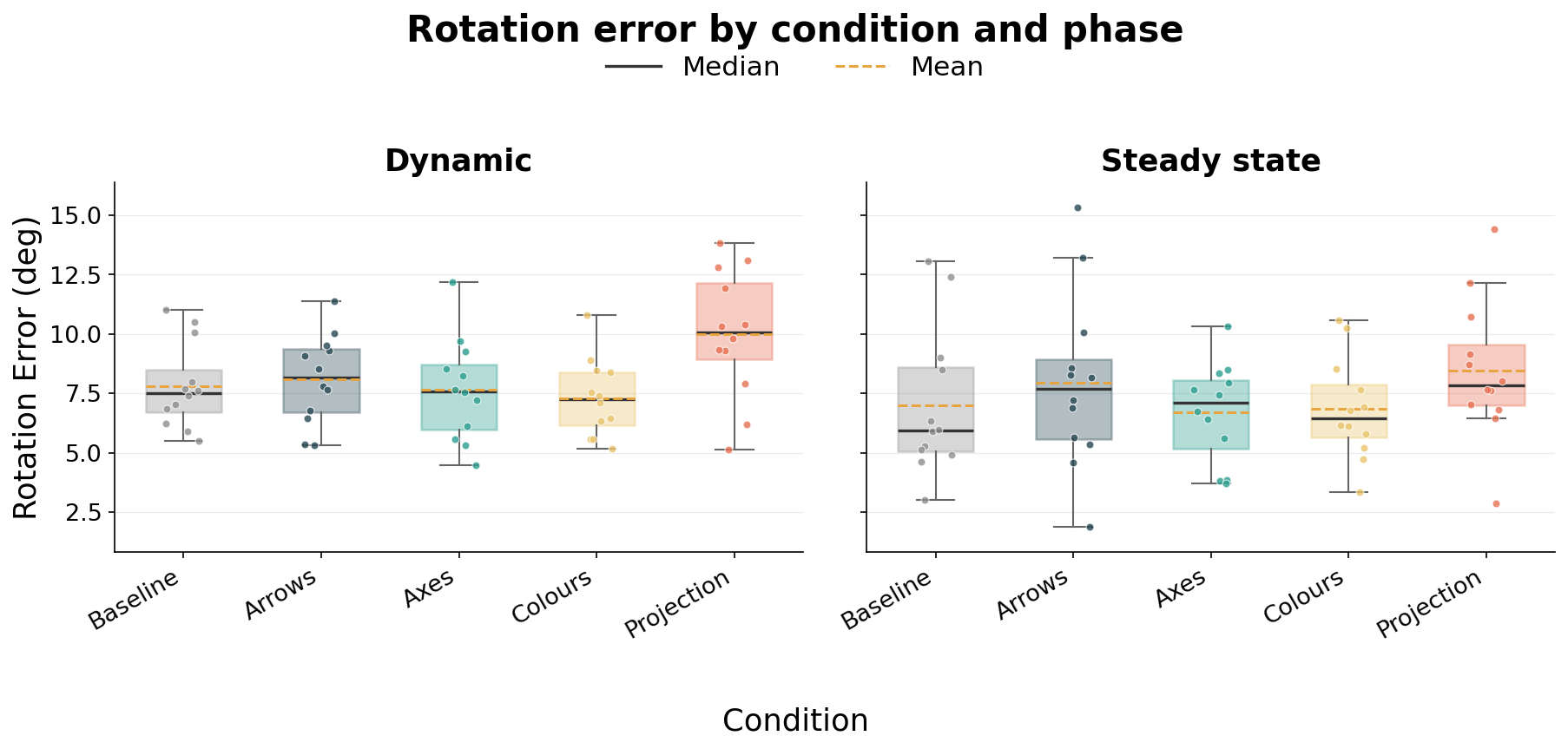}
    \caption{Rotational error by condition for the dynamic and steady-state phases. Each point represents one participant's mean error. Boxes show the interquartile range.}
    \label{fig:rot}
\end{figure}

Rotational error is shown in Figure~\ref{fig:rot} and Table~\ref{tab:rot_errors}. Projection significantly increased rotational error relative to Baseline in the dynamic phase ($p = 0.010$, $r_{\mathrm{rb}} = +0.92$). A similar effect was seen in the steady-state phase, but it did not reach significance ($p = 0.21$, $r_{\mathrm{rb}} = +0.64$). Axes and Colours achieved the lowest mean rotational error in both phases, though the differences were small and did not reach significance. Colours in the steady-state phase produced lower mean rotational error than Baseline, yet the effect size was positive ($r_{\mathrm{rb}} = +0.28$), with most participants producing higher rotational error than in the Baseline condition, while a few showed larger improvements.

\begin{table}[h!]
\centering\scriptsize
\renewcommand{\cellalign}{c}
\setlength{\tabcolsep}{3pt}
\caption{Mean rotational errors ($\pm$ SD) for the five conditions, by phase, with Holm-corrected paired Wilcoxon $p$-values and rank-biserial correlations $r_{\mathrm{rb}}$ against Baseline. Positive $r_{\mathrm{rb}}$ indicates higher error than Baseline. Significant values ($p<0.05$) are bold.}
\label{tab:rot_errors}
\begin{tabular}{@{}cl|c|cc@{}}
\toprule
\textbf{Condition} & \textbf{Phase} 
& \makecell{\textbf{Avg. rot} \\[-3pt] \textbf{error ($^\circ$)}} 
& \multicolumn{2}{c}{\textbf{Overall vs Baseline}} \\[-2pt]
 & & & $p$ & $r_{\mathrm{rb}}$ \\
\midrule
\multirow{2}{*}{Baseline} & Dynamic & 7.8 $\pm$ 1.8& -- & -- \\
 & Steady-state& 7.0 $\pm$ 3.1& -- & -- \\
\midrule
\multirow{2}{*}{Arrows} & Dynamic & 8.1 $\pm$ 1.9& 1.00 & $+0.21$ \\
 & Steady-state& 7.9 $\pm$ 3.7& 1.00 & $+0.28$ \\
\midrule
\multirow{2}{*}{Axes} & Dynamic & 7.7 $\pm$ 2.2& 1.00 & $-0.10$ \\
 & Steady-state& 6.7 $\pm$ 2.1& 1.00 & $-0.15$ \\
\midrule
\multirow{2}{*}{Colours} & Dynamic & 7.3 $\pm$ 1.6& 1.00 & $-0.28$ \\
 & Steady-state& 6.8 $\pm$ 2.1& 1.00 & $+0.28$ \\
\midrule
\multirow{2}{*}{\makecell[l]{Projection}} & Dynamic & 10.0 $\pm$ 2.7& \textbf{0.010} & $+0.92$ \\
 & Steady-state& 8.5 $\pm$ 3.0& 0.21 & $+0.64$ \\
\bottomrule
\end{tabular}
\end{table}

\begin{table}[h!]
\centering\scriptsize
\setlength{\tabcolsep}{3pt}
\caption{Alignment success rate by condition, with Holm-corrected paired Wilcoxon $p$-values and rank-biserial correlations $r_{\mathrm{rb}}$ against Baseline. Positive  $r_{\mathrm{rb}}$ indicates a higher success rate than Baseline and therefore better performance. Significant values ($p<0.05$) are bold.}
\label{tab:alignment}
\begin{tabular}{@{}lcccc@{}}
\toprule
\textbf{Condition}& \makecell{\textbf{Success} \\ \textbf{rate (\%)}} & \makecell{\textbf{Diff. vs} \\ \textbf{Baseline}} & \textbf{$p$} & \textbf{$r_{\mathrm{rb}}$} \\
\midrule
Baseline   & 15.0 & --    & --    & -- \\
Arrows     & 47.5 & +32.5 & \textbf{0.008}& $+1.00$ \\
Axes       & 37.5 & +22.5 & \textbf{0.047}& $+1.00$ \\
Colours    & 25.0 & +10.0 & 0.570& $+0.47$ \\
Projection & 20.0 & +5.0  & 0.570& $+0.36$ \\
\bottomrule
\end{tabular}
\end{table}

Success rate is shown in Table~\ref{tab:alignment}. Only 15.0\% of Baseline attempts reached a successful hold. Arrows raised this to 47.5\% ($p = 0.008$) and Axes to 37.5\% ($p = 0.047$), both significant improvements over Baseline. Colours (25.0\%) and Projection (20.0\%) both improved on Baseline, though not significantly, with moderate effect sizes ($r_{rb} = +0.47$ and $+0.36$, respectively). Notably, neither Arrows nor Axes reduced success for any participant, with rank-biserial correlations of $+1.00$ in both cases, indicating that every participant either improved or remained unchanged.

\begin{figure}[h!]
    \centering
    \includegraphics[width=0.75\linewidth]{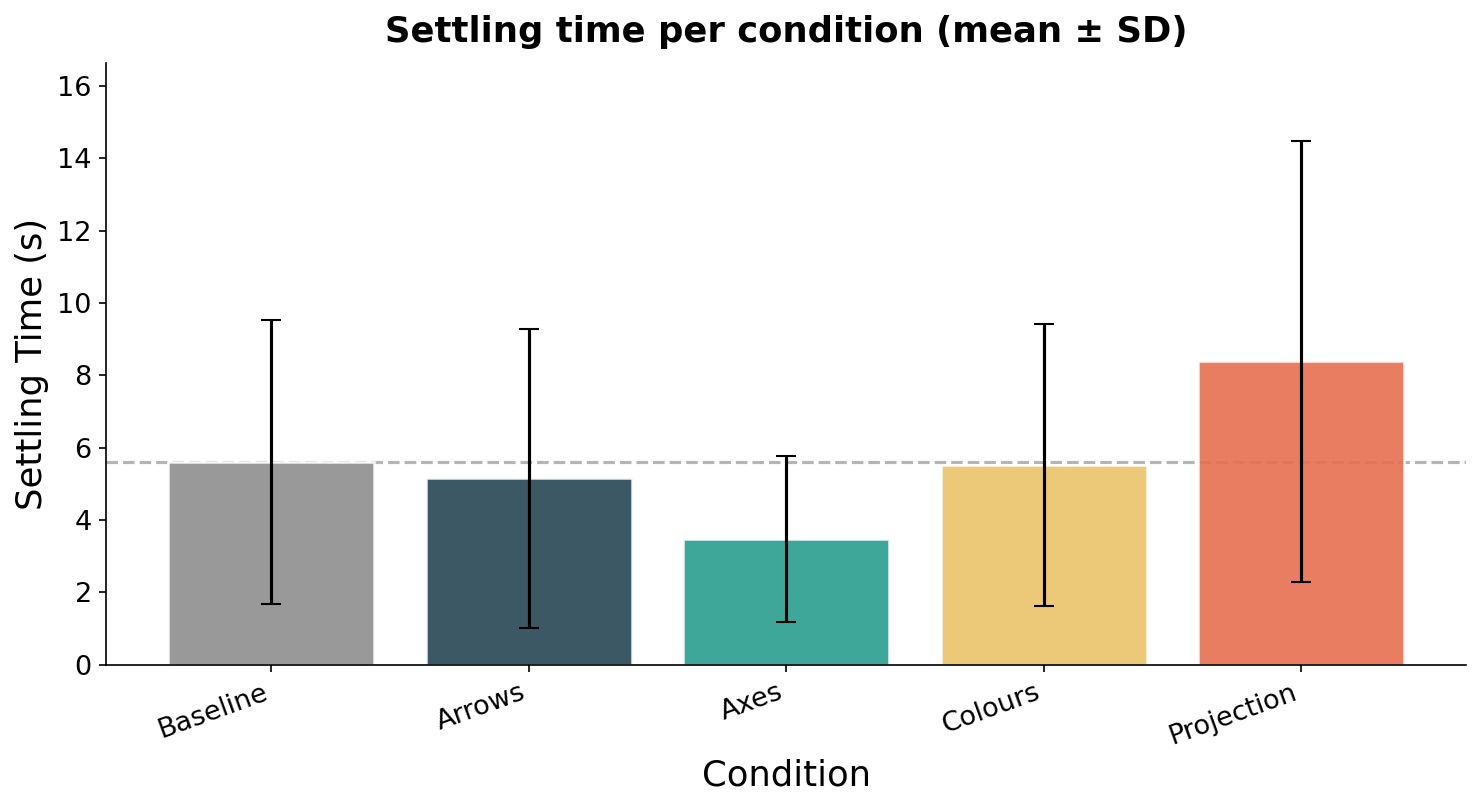}
    \caption{Mean settling time by condition among successful holds. Error bars show SD.}
    \label{fig:settling_time}
\end{figure}

Settling time by condition is shown in Figure~\ref{fig:settling_time}. Among successful holds, Axes produced the fastest mean settling time ($3.5 \pm 2.3$~s), followed by Arrows ($5.2 \pm 4.1$~s) and Colours ($5.5 \pm 3.9$~s), both close to Baseline ($5.6 \pm 3.9$~s). Projection produced the slowest mean settling time ($8.4 \pm 6.1$~s) and the greatest variability of any condition.

\begin{figure}[h!]
    \centering
    \includegraphics[width=1.0\linewidth]{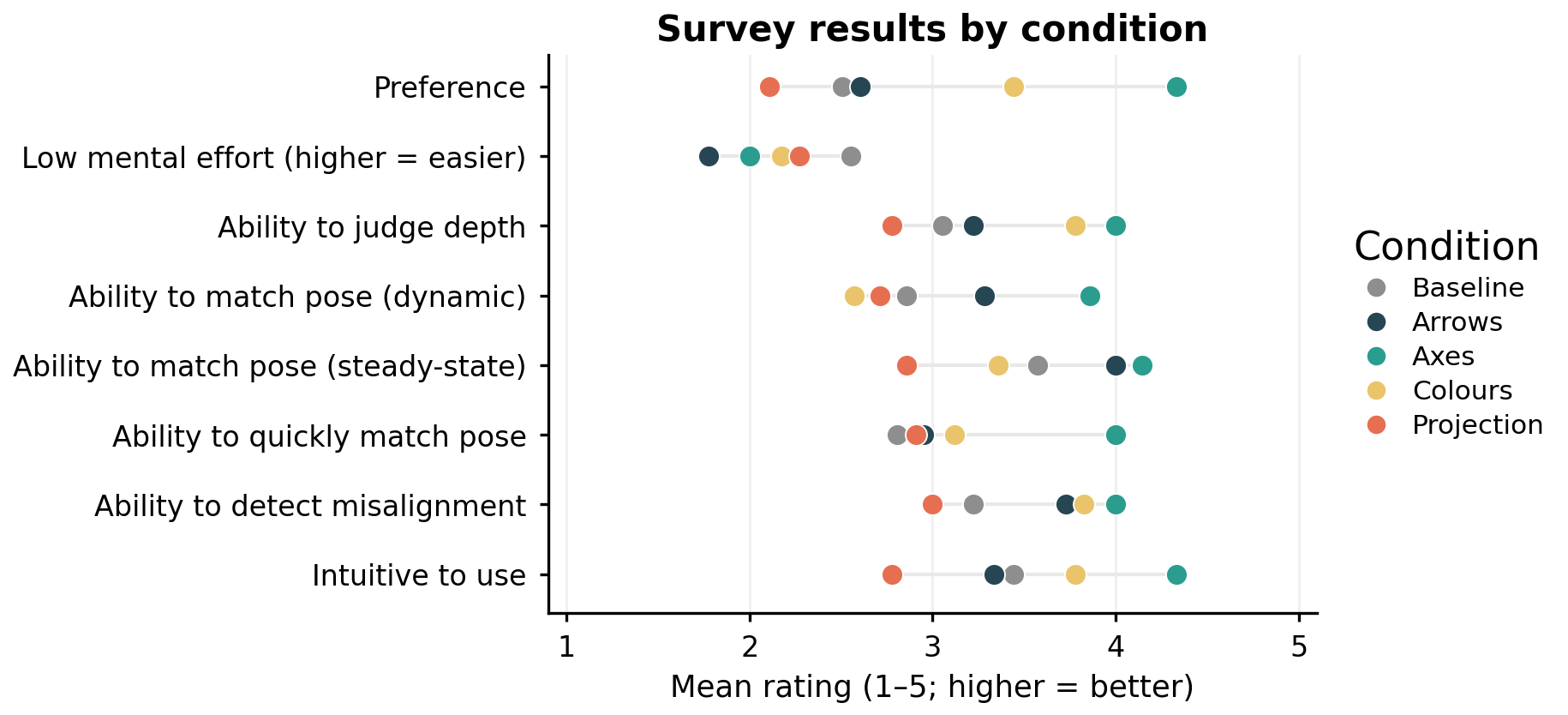}
    \caption{Mean survey ratings by condition across preference ranking and all Likert items (1--5, higher = better).}
    \label{fig:fig_cleveland_by_method}
\end{figure}

Participant survey results are shown in Figure~\ref{fig:fig_cleveland_by_method}. Axes was the highest-rated condition overall, with the highest mean preference score (4.3) and the highest rating on almost every individual Likert item, except low mental effort, where Baseline was rated highest and Arrows lowest.

\subsection{Study 2: Mobile vs.\ Magic Leap Comparison}
This section reports positional error, rotational error, active task duration, usability and image quality for the four AR guidance conditions evaluated in this work: Mobile R-V, Mobile V-V Feed, Mobile V-V No Feed, and Magic Leap. Results are presented as comparisons between each mobile configuration and the Magic Leap condition.

\begin{table}[h!]
\centering
\caption{Overall Study 2 results (mean $\pm$ SD). Part A includes all participants per condition ($n = 12$, Magic Leap and Mobile R-V from Day 1, V-V configurations from Day 2). Part B includes the eight participants who completed both sessions. Lines below each mobile configuration give the Holm-corrected paired Wilcoxon $p$-value and $r_{\mathrm{rb}}$ against Magic Leap, computed within that part. Positive $r_{\mathrm{rb}}$ indicates worse mobile performance (higher error or duration, or lower image quality). Significant values ($p < 0.05$) are bold.}
\label{tab:study2_overall}
\small
\renewcommand{\arraystretch}{1.25}
\setlength{\tabcolsep}{3pt}
\begin{tabular}{l|cc|cc|c|c}
\hline
 & \multicolumn{2}{c|}{\textbf{Position (mm)}} & \multicolumn{2}{c|}{\textbf{Rotation ($^\circ$)}}
 & \multirow{2}{*}{\makecell{\textbf{Duration}\\\textbf{(s)}}}
 & \multirow{2}{*}{\makecell{\textbf{Image quality}\\\textbf{(0--5)}}} \\
\cline{2-5}
\textbf{Condition} & Uncentered & Centered & Uncentered & Centered & & \\
\hline
\multicolumn{7}{l}{\textit{Part A. All participants per condition ($n = 12$)}} \\
\hline
\textbf{Magic Leap}
  & $22.9 \pm 9.5$ & $14.6 \pm 7.1$ & $11.0 \pm 3.4$ & $8.0 \pm 2.1$
  & $145.5 \pm 67.5$ & $3.8 \pm 0.3$ \\[8pt]
\textbf{Mobile R-V}
  & $28.2 \pm 8.4$ & $16.7 \pm 4.3$ & $10.6 \pm 3.0$ & $8.1 \pm 2.2$
  & $228.7 \pm 122.6$ & $3.5 \pm 0.5$ \\
  & \pv{0.353} & \pv{0.467} & \pv{1.000} & \pv{1.000} & \pv{0.103} & \pv{\textbf{0.047}} \\
  & \rb{+0.46} & \rb{+0.41} & \rb{-0.08} & \rb{+0.10} & \rb{+0.69} & \rb{+0.84} \\[8pt]
\textbf{Mobile V-V}
  & $16.3 \pm 5.0$ & $10.6 \pm 2.4$ & $11.4 \pm 2.5$ & $8.5 \pm 1.5$
  & $117.9 \pm 50.7$ & $4.1 \pm 0.3$ \\
\textbf{Feed} & & & & & & \\[8pt]
\textbf{Mobile V-V}
  & $18.2 \pm 7.8$ & $11.4 \pm 2.0$ & $10.4 \pm 2.2$ & $7.9 \pm 1.7$
  & $120.9 \pm 55.6$ & $4.3 \pm 0.3$ \\
\textbf{No Feed} & & & & & & \\
\hline
\multicolumn{7}{l}{\textit{Part B. Participants who completed both sessions ($n = 8$)}} \\
\hline
\textbf{Magic Leap}
  & $22.5 \pm 8.2$ & $15.5 \pm 8.5$ & $11.3 \pm 3.7$ & $8.3 \pm 2.3$
  & $144.7 \pm 77.3$ & $3.8 \pm 0.4$ \\[8pt]
\textbf{Mobile R-V}
  & $27.9 \pm 8.9$ & $16.5 \pm 4.3$ & $11.4 \pm 3.5$ & $8.6 \pm 2.5$
  & $254.9 \pm 143.1$ & $3.5 \pm 0.6$ \\[8pt]
\textbf{Mobile V-V}
  & $13.8 \pm 2.8$ & $9.8 \pm 1.5$ & $10.9 \pm 2.3$ & $8.1 \pm 1.5$
  & $108.6 \pm 60.0$ & $4.0 \pm 0.3$ \\
\textbf{Feed}
  & \pv{\mathbf{0.023}} & \pv{0.117} & \pv{1.000} & \pv{1.000} & \pv{0.297} & \pv{0.125} \\
  & \rb{-1.00} & \rb{-0.83} & \rb{-0.11} & \rb{0.00} & \rb{-0.61} & \rb{-0.71} \\[8pt]
\textbf{Mobile V-V}
  & $17.6 \pm 7.0$ & $11.5 \pm 2.2$ & $10.2 \pm 2.6$ & $7.8 \pm 1.9$
  & $123.2 \pm 66.1$ & $4.3 \pm 0.3$ \\
\textbf{No Feed}
  & \pv{0.353} & \pv{0.467} & \pv{1.000} & \pv{1.000} & \pv{0.297} & \pv{0.063} \\
  & \rb{-0.44} & \rb{-0.33} & \rb{-0.39} & \rb{-0.11} & \rb{-0.61} & \rb{-0.93} \\
\hline
\end{tabular}
\end{table}

\begin{figure}[!h]
    \centering
    \includegraphics[width=1.0\linewidth]{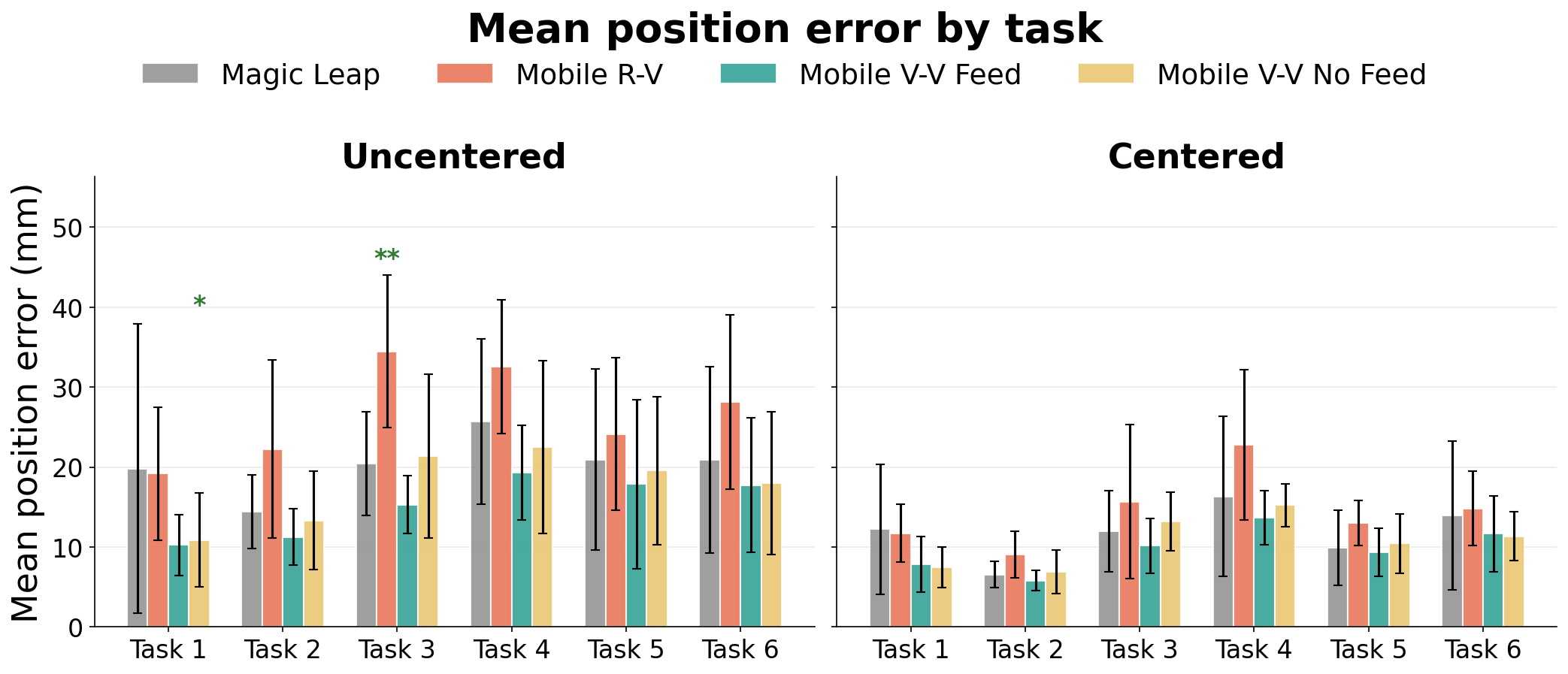}
    \caption{Mean positional error by condition and scanning task, uncentered (left) and centered (right). Error bars denote SD. Bars show all participants per condition. Significance markers for the V-V configurations use $n = 8$. Significant differences from Magic Leap (Holm-corrected) are marked with * for $p<0.05$ and ** for $p<0.01$; exact $p$ and $r_{\mathrm{rb}}$ values are reported in Supplementary Table S1.}
    \label{fig:study2_pos_task}
\end{figure}

Figure~\ref{fig:study2_pos_task} reports positional error for each of the six scanning tasks (Section~\ref{study2_procedure}), with overall means reported in Table~\ref{tab:study2_overall} and exact per-task values given in Supplementary Table~S4. Mean overall positional error was highest for Mobile R-V, followed by Magic Leap, with Mobile V-V No Feed and Feed configurations producing the lowest error. Centering reduced positional error across all four conditions, with the same ordering preserved.

Mobile R-V did not differ significantly from Magic Leap overall ($p = 0.353$, $r_{\mathrm{rb}}=+0.46$; Supplementary Table~S1), but had significantly higher error on Task 3 (34.5~mm vs.\ 20.4~mm, $p = 0.009$, $r_{\mathrm{rb}}=+0.95$). Task 2 also showed a higher error for Mobile R-V (22.2~mm vs.\ 14.4~mm, $p = 0.081$, $r_{\mathrm{rb}}=+0.77$), but this did not reach significance after Holm correction.

Among the eight participants who completed both sessions, Mobile V-V Feed had significantly lower overall error than Magic Leap (13.8~mm vs.\ 22.5~mm, $p = 0.023$, $r_{\mathrm{rb}}=-1.00$), with every one of these participants showing lower error, although no individual task reached significance. Mobile V-V No Feed did not differ significantly from Magic Leap overall ($p = 0.353$, $r_{\mathrm{rb}}=-0.44$), but had significantly lower error on Task 1 (9.0~mm vs.\ 15.9~mm, $p = 0.047$, $r_{\mathrm{rb}}=-1.00$), just reaching significance. Both V-V configurations had comparable or slightly lower mean error than Magic Leap across all tasks. 

After centering, the Mobile R-V Task 3 difference was no longer significant ($p = 1.000$, $r_{\mathrm{rb}}=+0.28$), and neither was the Mobile V-V Feed overall difference (9.8~mm vs.\ 15.5~mm, $p = 0.117$, $r_{\mathrm{rb}} = -0.83$). No comparison, for any configuration or task, reached significance after centering.

\begin{figure}[H]
    \centering
    \includegraphics[width=1.0\linewidth]{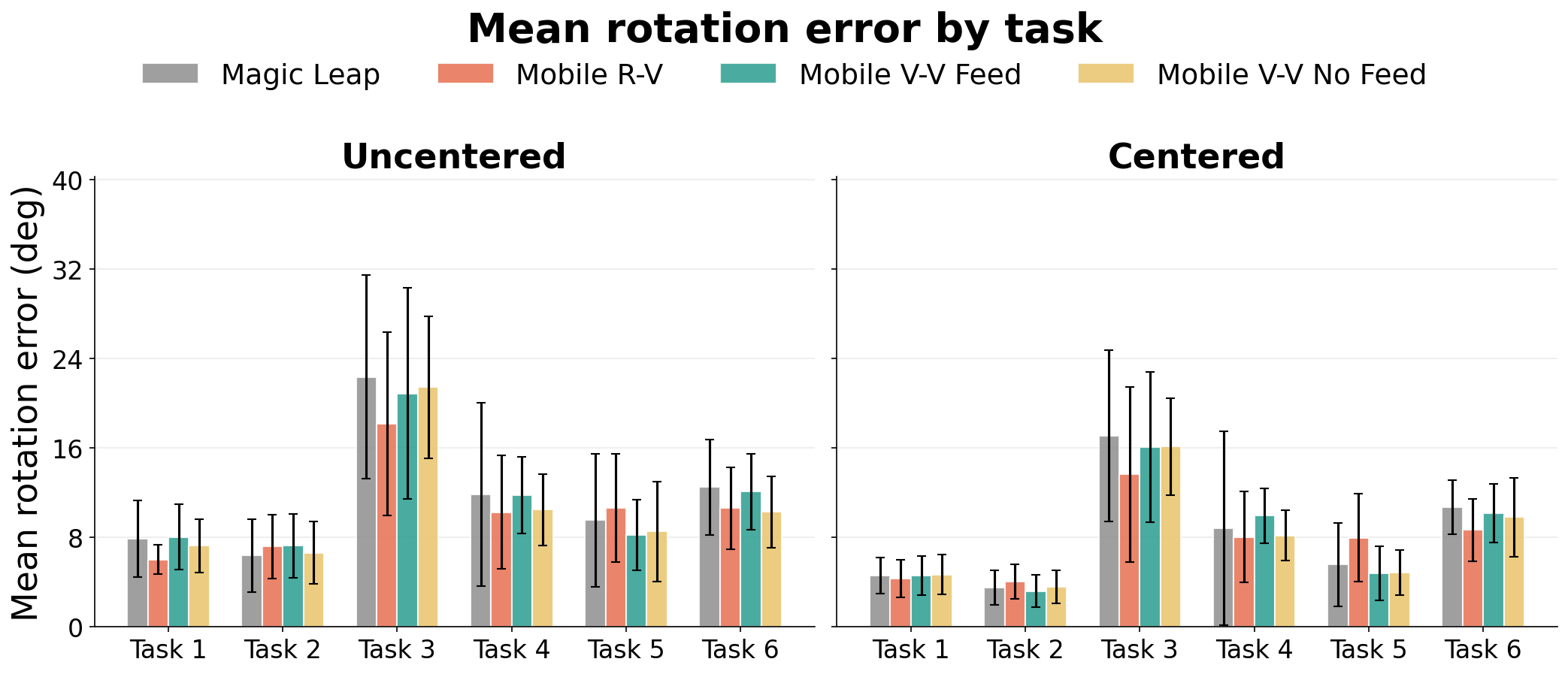}
    \caption{Mean rotational error by condition and scanning task, uncentered (left) and centered (right). Error bars denote SD. Bars show all participants per condition. Significance markers for the V-V configurations use $n = 8$. Exact $p$ and $r_{\mathrm{rb}}$ values are reported in Supplementary Table S2.}
    \label{fig:study2_rot_task}
\end{figure}

Figure~\ref{fig:study2_rot_task} reports rotational error for each scanning task, with overall means in Table~\ref{tab:study2_overall} and exact values given in Supplementary Table~S4. Overall uncentered rotational error was similar across conditions, and none of the three mobile configurations differed significantly from the Magic Leap, either overall or by task, before or after centering (Supplementary Table~S2). Differences varied in direction across tasks, with no configuration consistently higher or lower. Task 3 produced the largest error by a wide margin for all four conditions (18.1$^{\circ}$--22.3$^{\circ}$).

\begin{figure}[H]
    \centering
    \includegraphics[width=1.0\linewidth]{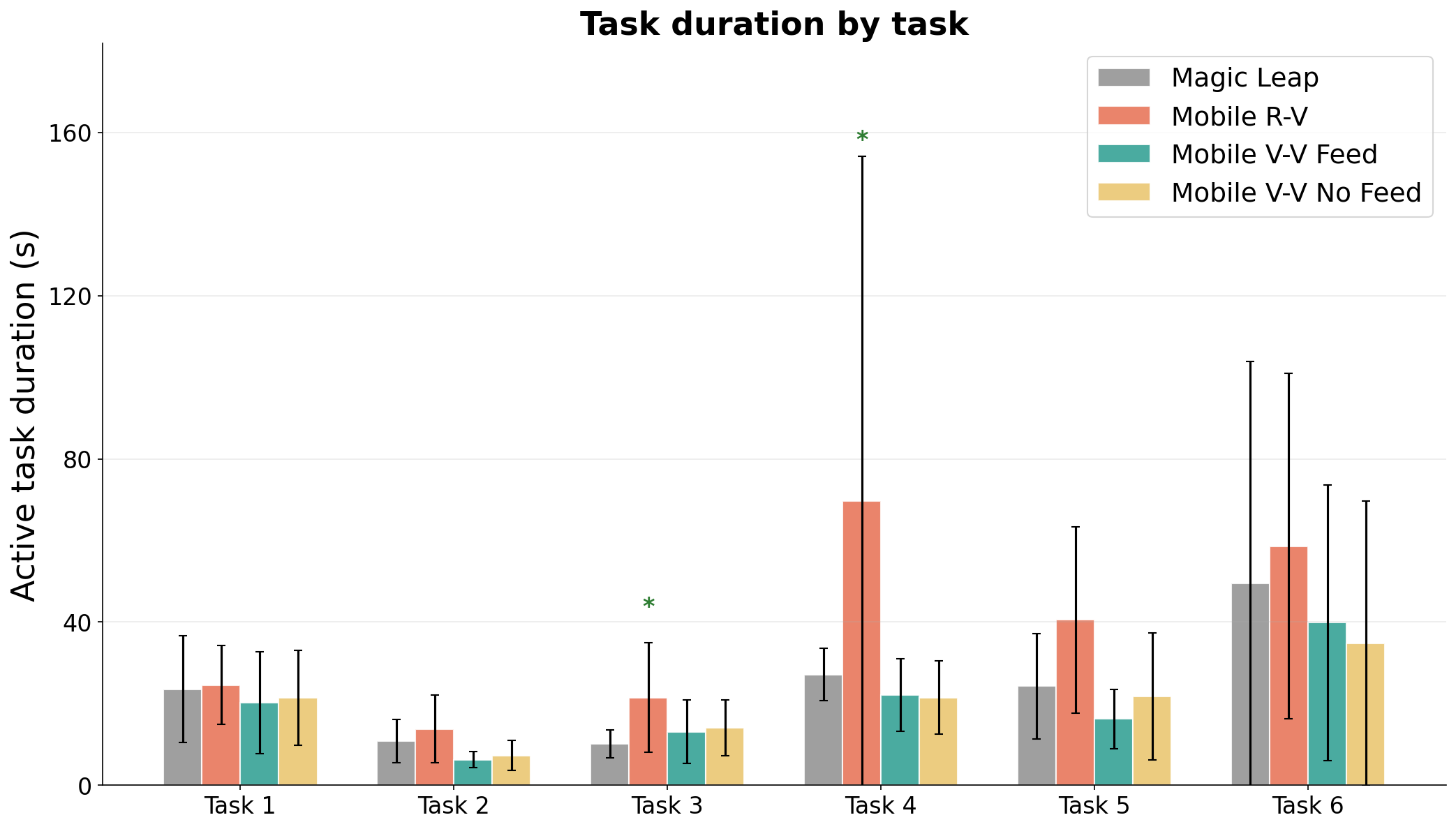}
    \caption{Active task duration by condition and scanning task. Error bars denote SD. Bars show all participants per condition. Significance markers for the V-V configurations use $n = 8$. Significant differences from Magic Leap (Holm-corrected, $p<0.05$) are marked with *; exact $p$ and $r_{\mathrm{rb}}$ values are reported in Supplementary Table S3.}
    \label{fig:study2_dur_task}
\end{figure}

Figure~\ref{fig:study2_dur_task} reports active task duration for each scanning task, with overall means reported in Table~\ref{tab:study2_overall} and exact values given in Supplementary Table~S4. Magic Leap was faster overall than Mobile R-V ($145.5 \pm 67.5$~s vs.\ $228.7 \pm 122.6$~s), but slower than both Mobile V-V configurations among the eight participants who completed both sessions ($144.7 \pm 77.3$~s vs.\ $108.6 \pm 60.0$~s for Feed and $123.2 \pm 66.1$~s for No Feed).

Mobile R-V took significantly longer than Magic Leap on Task 3 ($21.5$~s vs.\ $10.2$~s, $p = 0.015$, $r_{\mathrm{rb}}=+0.92$) and Task 4 ($69.7$~s vs.\ $27.2$~s, $p = 0.034$, $r_{\mathrm{rb}}=+0.85$), although the overall difference was not significant ($p = 0.103$, $r_{\mathrm{rb}}=+0.69$). Both Mobile V-V configurations were faster than Magic Leap on all tasks except for Task 3, but none differed significantly at the task level or overall. Despite the varied completion times, all participants successfully completed every examination in all conditions.

Table~\ref{tab:survey_summary} reports SUS, NASA-TLX, AR guidance quality, and preference rank across all four conditions. Mobile R-V received the lowest usability score of the four conditions (SUS: $62.7 \pm 14.0$), while Magic Leap and both V-V configurations scored higher and similarly to each other (80.0--80.9).

\begin{table}[H]
\centering\scriptsize
\setlength{\tabcolsep}{3pt}
\caption{Participant survey results across all four conditions (mean $\pm$ SD). Higher is better for SUS and AR guidance quality. Lower is better for NASA-TLX and Preference rank.  Preference rank includes only the eight participants who completed both sessions.}
\label{tab:survey_summary}
\begin{tabular}{lcccc}
\toprule
Condition & SUS (0--100) & NASA-TLX (0--100) & AR guidance quality (1--5) & Preference rank (1--4) \\
\midrule
ML & $80.0 \pm 10.1$ & $20.5 \pm 14.7$ & $4.2 \pm 0.4$ & $1.9 \pm 1.1$ \\
R-V & $62.7 \pm 14.0$ & $38.4 \pm 18.9$ & $3.7 \pm 0.9$ & $4.0 \pm 0.0$ \\
V-V Feed & $80.9 \pm 13.6$ & $26.8 \pm 15.3$ & $4.2 \pm 0.8$ & $2.0 \pm 0.8$ \\
V-V No Feed & $80.6 \pm 18.2$ & $26.8 \pm 18.4$ & $4.3 \pm 0.4$ & $2.1 \pm 0.7$ \\
\bottomrule
\end{tabular}
\end{table}

A similar pattern was observed for perceived workload. Mobile R-V also showed the highest perceived workload (NASA-TLX: 38.4), compared with 20.5 for Magic Leap and 26.8 for both V-V configurations. AR guidance quality ratings were high across all conditions (3.7--4.3 out of 5), though Mobile R-V was the lowest. Preference rank had Magic Leap ranked best on average ($1.9 \pm 1.1$), followed closely by V-V Feed ($2.0 \pm 0.8$) and V-V No Feed ($2.1 \pm 0.7$), and Mobile R-V ranked worst and last by every participant ($4.0 \pm 0.0$).

Table~\ref{tab:study2_overall} reports mean image quality scores for each condition. Intra-rater reliability was high, with 75\% exact agreement, 100\% agreement within 1 point, and a quadratic-weighted kappa of 0.88. Mean scores ranged from 3.5 (Mobile R-V) to 4.3 (Mobile V-V No Feed). Mobile R-V scored significantly lower than Magic Leap ($p = 0.047$, $r_{\mathrm{rb}} = +0.84$), just reaching significance, while neither V-V configuration differed significantly from Magic Leap (Feed: $p = 0.125$, No Feed: $p = 0.063$).

\section{Discussion}\label{discussion}
\subsection{Study 1: Guidance Cue Comparison}
\label{discussion_study1}
\subsubsection{Cue Performance}
The Study 1 findings suggest that each cue affected different aspects of probe guidance. Colours provided the clearest benefit for positional accuracy, particularly along the depth axis, whereas Arrows and Axes were more effective at helping participants achieve the defined alignment criterion during the steady-state phase. Projection provided limited positional benefit and worsened rotational performance. 

Across all conditions, the Z-axis exhibited the largest positional error, consistent with the known difficulty of depth perception in handheld AR. This points to depth estimation as a primary limitation and highlights the importance of additional cues to compensate. For Arrows, Axes, and Colours, X-axis and Y-axis error remained comparable to Baseline, suggesting that improvements in overall positional accuracy were primarily driven by reductions in depth error. Projection was an exception, showing increased Y-axis error in both phases~(Table~\ref{tab:study1_pos_errors}).

Colours was the only cue to significantly improve positional accuracy relative to Baseline in both phases, driven primarily by reduced depth-axis error (37\% and 54\% reductions in the dynamic and steady-state phases, respectively). The greater reduction during the steady-state phase may reflect the additional time available to interpret the colour feedback and refine position, whereas dynamic tracking required participants to adjust continuously. Colours also produced slightly lower mean rotational error than Baseline, although neither difference was statistically significant and most participants had higher error with Colours in the steady-state phase. This is consistent with Colours conveying positional but not rotational error.

Arrows and Axes showed a different pattern, providing more benefit in achieving the alignment criterion than in reducing overall positional error. Both significantly improved alignment success without significantly reducing overall positional error, though both produced lower mean errors than Baseline, with moderate to large effect sizes. Given the small sample size, the study may not have had sufficient power to detect the positional differences. 

The lower mean positional error with Colours may seem inconsistent with its lower success rate. However, this likely reflects the distribution of errors around the 5~mm success threshold. Colours produced lower variability across participants~(Figure~\ref{fig:pos}), with many holds clustered near the threshold, whereas Arrows and Axes produced a wider distribution that resulted in more holds falling below 5~mm despite some larger errors. The transition towards green near the 5~mm threshold may have made these holds appear sufficiently aligned despite remaining outside the success criterion. In contrast, Arrows disappeared entirely when the probe was within 5~mm, directly indicating that the tolerance had been reached, while Axes provided a stable spatial reference to overlap. Survey comments also indicated some confusion regarding the direction represented by Colours, which may have contributed to oscillations around the target.

Settling time results were partially consistent with the success rate findings. Axes produced the fastest mean settling time, suggesting more efficient convergence when participants achieved a successful hold. Arrows achieved the highest success rate but had a settling time comparable to Baseline, indicating that while it increased the likelihood of successful alignment, it did not necessarily reduce the time required. No statistical comparisons were performed for settling time as the number of successful holds differed across conditions. 

Axes did not significantly reduce rotational error, potentially because rotational accuracy was already relatively stable under the Baseline condition and therefore may not be a major limiting factor for probe guidance. Arrows produced a small, non-significant increase in rotational error, potentially reflecting the attention division between the cue and probe itself, consistent with previous findings that excessive visual feedback can increase cognitive load~\citep{black2023humanasrobot}.

Projection was the only cue to degrade performance relative to Baseline, increasing rotational error during both phases and significantly during dynamic tracking. Although depth error showed a modest improvement, overall positional accuracy and alignment success were largely unchanged. Since the reticle was spatially separated from the probe, participants may have needed to divide attention between the cue and probe, contributing to the increased rotational error. This contrasts with Andersen and Popescu~\citeyearpar{andersen2020ar}, whose projection-based cue improved rotational accuracy, though their reticle incorporated explicit rotational information and was rendered directly on the object, which may explain the differing findings. This motivates future work on a projection cue that remains spatially coupled to the probe while conveying rotation explicitly.

Although positional errors were generally lower during the steady-state phase, the relative pattern of cue performance remained consistent across the two phases. Statistical significance varied by phase, but the direction of cue benefits remained similar, suggesting that probe motion influenced the magnitude of the cue effects rather than their direction.

Subjective preference did not fully correspond to objective performance. Axes was the most preferred cue despite not producing the lowest positional errors or highest alignment success rate, whereas Colours produced the lowest positional and depth errors but was not the most preferred. Arrows achieved the highest alignment success rate but was described as mentally fatiguing and difficult to track. These findings suggest that perceived ease of use does not necessarily correspond to measurable performance improvements, and motivated the combination of the Colours and Axes cues for their complementary strengths in the obstetric US study (Section~\ref{ML_vs_mobile}). Arrows were not retained because participants found them mentally demanding and confusing, whereas Axes provided a comparable alignment reference with lower reported cognitive demand and an additional target pose reference.

\subsubsection{Limitations}\label{study1_limitations}
Several limitations should be considered. First, the probe trajectories did not correspond to clinical US scans but instead consisted of scan-like movements designed to test alignment during stationary and dynamic conditions. The trials were also performed on a compliant material rather than actual human tissue, which does not fully capture clinical conditions such as tissue deformation. Although condition and trajectory order were randomized, individual trajectories may have differed in difficulty, introducing the possibility that certain trajectory-cue combinations influenced performance. The mobile system also introduced ergonomic limitations. The smartphone display provided a smaller field of view than headset-based systems, requiring participants to balance visibility of the workspace with a comfortable viewing distance and arm position. These constraints may have influenced both usability and task performance. Finally, the study included only twelve participants, limiting statistical power to detect smaller differences between cues.

Study 1 was also limited because the virtual probes and guidance cues were parented to the mobile camera rather than the Unity world frame. This approach was shaped by earlier testing, where world-frame placement caused the virtual objects to drift relative to the physical scene when AR tracking lost or lacked sufficient visual features to maintain a stable world frame. Anchoring the overlay to an AR-detected surface did not fully resolve the issue, as the overlay could still drift when the plane was lost or the reconstructed surface was inaccurate. Anchoring to an ArUco marker provided greater stability, but the camera's autofocus tended to favour the nearby probe over the more distant marker. The resulting blur degraded marker pose estimates and reintroduced drift. Parenting the overlay to the camera avoided these issues and kept it stable in the user's view. This was acceptable for Study 1 since the phone remained stationary in a holder throughout each trial, and these tracking limitations may be specific to the Android device used in this study.

\subsection{Study 2: Mobile vs.\ Magic Leap Comparison}
\subsubsection{Alignment Accuracy and Task Duration}
None of the mobile configurations had significantly higher overall positional or rotational error than the Magic Leap system, and Mobile V-V Feed had significantly lower overall positional error. Differences between conditions appeared mainly in positional error, while rotational error remained similar. 

Both uncentered and centered positional error were considered when comparing the conditions. The centered positional error is particularly relevant to this probe guidance task, as the follower was not being guided toward a known absolute probe position. Instead, the expert continuously viewed the follower's US stream and provided guidance based on the resulting image, meaning that guidance depended more on the relative than the absolute position of the probes. Removing a constant offset provides a better indication of how closely the follower reproduced the expert's probe movements. 

Mobile R-V's uncentered positional error was higher than Magic Leap's, with the largest difference on Task 3, the only task where it was significantly higher ($p = 0.009$, $r_{\mathrm{rb}} = +0.95$). This difference was reduced after centering, suggesting it partly reflects a consistent task-level offset rather than differences in how closely participants followed the expert's probe motion. After centering, Mobile R-V performed similarly to Magic Leap, with a slightly higher error that was not significantly different.

The consistent offset could have several sources. Small calibration or registration error in the ArUco-based tracking could introduce a consistent offset between the estimated and true probe poses. In the Mobile R-V configuration, this could shift the Colours and Axes cues slightly away from the physical probe, creating a mismatch between the visual feedback and the actual R-V alignment. 

A similar issue could arise from registration of the virtual phantom to the physical phantom. If the virtual surface were slightly offset from the physical surface, the expert and follower could each maintain probe contact relative to their respective phantom surfaces while having a consistent offset between their tracked probe poses. Centering removes these consistent offsets, but cannot determine their source or account for movements unrelated to probe alignment. For example, participants occasionally repositioned themselves or reached for US gel, and these movements could have contributed to the measured positional error.

Both V-V configurations, in contrast, had lower overall uncentered and centered positional error than Magic Leap. Mobile V-V Feed had lower uncentered error on every task and significantly lower overall error, with all eight participants performing better than with Magic Leap. Mobile V-V No Feed had lower uncentered error on five of six tasks, significantly so on Task 1, but lower centered error on only three. Among the eight participants who completed both sessions, overall centered error was 5.7~mm lower for Mobile V-V Feed and 4.0~mm lower for Mobile V-V No Feed than for Magic Leap, although neither difference was significant after centering.

This suggests that the V-V representation may support accurate alignment, with both probes represented in the virtual scene potentially reducing visual ambiguity. This advantage may also partly reflect how error was measured. In all configurations, error was measured between the expert probe and the tracked pose of the physical follower probe. In the V-V configurations, participants aligned a virtual follower probe drawn at this tracked pose, so they were aligning the same pose used to measure error. Any difference between the tracked and true pose of the physical probe, such as from the ArUco calibration error described above, would not appear in the measured error. In Mobile R-V, participants aligned the physical probe directly, so any such difference could contribute to the measured error. Part of the lower V-V error may therefore reflect this difference in measurement.

The two V-V configurations showed similar performance overall. Removing the camera feed and scaling down the scene in Mobile V-V No Feed did not produce a clear improvement in performance relative to the Mobile V-V Feed configuration.

Rotational error showed much less separation between the conditions, with none of the mobile configurations differing significantly from Magic Leap. Centering had little effect on the comparison, suggesting there was no systematic rotational offset between conditions, and rotational error was not consistently higher or lower for any mobile configuration across tasks.

Task duration followed a similar pattern to positional error. Mobile R-V took longer than Magic Leap overall, although not significantly, and was significantly slower on Tasks 3 and 4. Both V-V configurations were faster than Magic Leap on five of six tasks and overall, although none of these differences were significant. 

\subsubsection{Usability, Preference, and Image Quality}
Survey results were broadly consistent with the objective performance measures. Mobile R-V received the lowest usability scores, highest perceived workload, and was ranked last by every participant. Both V-V configurations scored close to Magic Leap on usability and workload. AR guidance quality ratings were comparably high across all four conditions, with Mobile R-V rated only slightly lower, suggesting the guidance cues themselves were not the primary source of the usability and workload differences.

The two main distinctions between the Mobile R-V and the V-V configurations were that the phone was handheld versus mounted, and that alignment was matching a real probe to a virtual target versus matching two virtual probes within the same scene. Both likely contributed to the improved usability and reduced workload with the V-V configurations. Participant comments supported this. For Mobile R-V, several participants described difficulties holding the phone and probe simultaneously. They noted that keeping the probe within the camera's field of view required holding the phone close to the face, increasing both mental and physical effort, while also making it difficult to reposition their grip on the probe.

However, field of view and positioning constraints were not fully resolved by mounting the phone in the V-V configurations. Participants using the mounted setup still reported awkward arm positioning to see the screen comfortably, describing needing to lean back or hold their arm at a strained distance to keep the probe in view while viewing the screen comfortably. This suggests that mounting the phone freed both hands for probe manipulation, but did not fully eliminate the underlying field of view limitation of the smartphone.

Few of the eight participants who used both R-V and V-V commented on alignment comparison directly, but those who did found V-V easier, consistent with its lower positional error. Between the V-V configurations, the majority of participants preferred the camera feed, mentioning easier depth judgment, more accurate probe alignment, and better spatial awareness. A few participants preferred no camera feed for its wider field of view and reduced distraction from the live video. This points to a modest preference for retaining the camera feed, consistent with the similar performance and usability observed between the two configurations.

Image quality ratings further supported the feasibility of the Mobile V-V configurations. Mobile R-V produced significantly lower image quality than Magic Leap, while neither V-V configuration differed significantly from Magic Leap. Both Mobile V-V configurations had higher mean scores than Magic Leap and Mobile R-V, consistent with their lower mean positional error than both conditions.

Across all conditions, every participant successfully completed all six tasks, so the differences between conditions reflect variation in alignment accuracy, completion time, usability, and image quality rather than an inability to perform and complete the examination. 

Although the study was not designed to test equivalence, the differences observed between the mobile configurations and Magic Leap were small. Mobile R-V, the lowest-performing configuration, had a mean centered positional error about 2~mm higher than Magic Leap and a rotational error within 0.5$^\circ$. It did, however, take longer on some tasks, receive lower usability ratings, and produce significantly lower image quality ratings, suggesting that this configuration setup may have limited performance. The V-V configurations addressed these issues, producing lower positional error than Magic Leap, significantly so overall for Mobile V-V Feed, along with a consistent trend toward shorter task durations, and comparable usability and diagnostic image quality. V-V alignment may therefore be a useful guidance approach independent of platform and could be worth exploring within the Magic Leap system as well. Although the V-V advantage may partly reflect the session order and other limitations discussed below, even a cautious interpretation suggests that the V-V mobile configurations performed at least similarly to the headset.

\subsubsection{Limitations}
Several limitations should be considered in interpreting these results. While the Mobile R-V comparisons used all twelve Day 1 participants, the V-V comparisons with Magic Leap were based on only eight participants, because only those who completed both study sessions were included in the paired analysis. This limits power to detect smaller differences, since with $n = 8$ a comparison could only reach significance after correction when nearly all participants differed in the same direction. The V-V configurations were also always evaluated in a second session, after participants had already used the Mobile R-V configuration and the Magic Leap system. Because task familiarity, phone mounting, and alignment approach (R-V or V-V) all changed together between sessions, the present study cannot isolate their individual contributions to the observed differences. The same expert sonographer guided both sessions, so their familiarity with the task set may also have contributed to the shorter durations recorded in the second session.

Another limitation is that the Magic Leap system used a default rendering without the silhouette outline, Colours, or Axes cues used in the mobile configurations, as these cues were developed as part of this work. Differences between the mobile configurations and Magic Leap may partly reflect the added guidance cues rather than the display platform itself, making it difficult to isolate the effect of the platform. Nevertheless, the comparison supports the feasibility of the mobile system as an alternative to the original Magic Leap system.

Pose errors were computed from each platform's own tracking, using the headset's built-in ArUco tracking for Magic Leap and the custom ArUco pipeline for the mobile configurations. Any difference in accuracy between these two tracking systems was not quantified, but could have contributed to the measured differences between platforms. If the mobile tracking was less accurate, the ArUco-based pipeline could be revisited and refined to improve registration accuracy, potentially improving the alignment between the visual cues and the physical probe in the R-V configuration. With the existing tracking and experimental setup, the phone's position was constrained because the ArUco markers had to stay in view, requiring the device to remain farther back to track the probe across the full workspace. This limited some of the hypothesized benefits of Mobile V-V No Feed, where the scaled-down scene was intended to allow the device to be placed closer to the workspace. Future work exploring alternative tracking methods could reduce this constraint, allowing the phone to be placed closer, which may further improve ergonomics and performance.

These results were also obtained using an obstetric phantom in a controlled setting rather than a real patient. Performance may differ under clinical conditions, where factors such as patient movement and the need to interpret anatomy introduce additional challenges. Future work could evaluate the system with a larger participant sample and more realistic US scanning tasks, including real patient imaging, to better assess its performance in clinical settings and provide greater power to determine whether the V-V advantage over Magic Leap is reliable. All images were also rated by a single radiologist, so intra-rater reliability could be assessed but inter-rater reliability could not.

\section{Conclusion}
This work presented a smartphone-based implementation of Human Teleoperation for tele-ultrasound guidance and investigated how visual cues and virtual probe representations support close-range 6-DoF alignment on a handheld AR display in two user studies. The first study identified colour, directional arrows, and coordinate axes cues as useful for real-to-virtual probe alignment, with the colour cue producing the largest reduction in positional error and the directional arrows and coordinate axes producing the highest alignment success rates. The second study combined the colour and coordinate axes cues in three mobile configurations that compared aligning the physical follower probe with aligning a virtual representation of it, and evaluated them against the original mixed-reality headset during a simulated obstetric ultrasound scan. Overall, no mobile configuration performed significantly worse than the Magic Leap in overall positional or rotational error, or task duration. Compared to the Magic Leap, the mobile real-to-virtual configuration produced significantly lower image quality, and higher error and longer duration on some individual tasks, whereas the virtual-to-virtual configurations showed lower mean positional error than the Magic Leap and comparable usability ratings. All participants successfully acquired all images with each condition, showing that the mobile system could support the full scanning task, although the poorer performance of the real-to-virtual configuration suggests that the visual representation and physical interaction with the device are important design considerations for mobile AR guidance.

These findings suggest that a smartphone, particularly with a virtual-to-virtual configuration, could provide a lower-cost alternative to a mixed-reality headset for performing remote ultrasound using the Human Teleoperation system. Since smartphones are widely owned and readily available, they may provide a more accessible option for deployment in rural and remote settings where access to trained sonographers is limited. More broadly, the results indicate that aligning virtual representations rather than physical objects, supported by targeted depth cues, may benefit other handheld AR tasks requiring precise remote guidance. Future work should evaluate the system with a larger participant sample and real patient imaging.

\section*{Acknowledgements}
The authors would like to thank all participants for their time and contribution to this study.

\section*{Ethics statement}
This study was performed in line with the principles of the Declaration of Helsinki and approval of the University of British Columbia Behavioural Research Ethics Board (Approval No: H22-01195; June 20, 2022). All participants provided informed consent prior to participation. 

\section*{Author contributions}
CRediT: \textbf{Elizabeth Klosa}: Conceptualization, Data curation, Formal analysis, Investigation, Methodology, Project administration, Software, Validation, Visualization, Writing -- original draft; \textbf{Inzaghi Moniaga}: Conceptualization, Software; \textbf{Victoria Lessoway}: Investigation, Methodology; \textbf{Jessica Li}: Investigation; \textbf{David Black}: Conceptualization, Methodology, Resources, Software, Supervision, Writing -- review \& editing; \textbf{Septimiu Salcudean}: Conceptualization, Funding acquisition, Resources, Supervision, Writing -- review \& editing.

\section*{Disclosure statement}
No potential conflict of interest was reported by the author(s).

\section*{AI usage disclosure}
Generative AI tools, ChatGPT (GPT-5, OpenAI) and Claude (Opus 5.5, Anthropic), were used to assist with identifying relevant literature, language refinement, and structural editing during manuscript preparation. Google Gemini (3.1 Flash Image) was used to generate conceptual illustrations, as specified in the relevant figure captions. The authors checked the terms of use for each tool, reviewed and verified all content and take full responsibility for the originality, accuracy, and integrity of the manuscript, including the accuracy of all references.

\section*{Funding}
This work was supported by a Rogers-MITACS grant, the NSERC CGS M scholarship held by E. Klosa, the NSERC postdoctoral fellowship held by Dr. D. Black, and the UBC C.A. Laszlo Chair held by Prof. Salcudean.

\section*{Data Availability}
Data are available upon reasonable request. 

\bibliographystyle{apacite}
\bibliography{bibliography}

\clearpage
\section*{Supplementary Material}
\setcounter{table}{0}
\renewcommand{\thetable}{S\arabic{table}}

Tables~\ref{tab:S1} to \ref{tab:S4} report the task-level results for Study~2. 

\begin{table}[H]
\centering\small
\caption{Statistical comparisons for positional error, comparing each mobile configuration with Magic Leap. Cells report Holm-corrected $p$ values with matched-pairs rank-biserial correlations ($r_{\mathrm{rb}}$) in brackets. R-V vs ML uses $n = 12$, while Feed and No Feed vs ML use the $n = 8$ participants who completed both sessions. Positive $r_{\mathrm{rb}}$ indicates higher values for the mobile configuration. Significant values ($p < 0.05$) are bold.}
\label{tab:S1}
\begin{tabular}{llccc}
\toprule
Task & & R-V vs ML & Feed vs ML & No Feed vs ML \\
\midrule
\multirow{2}{*}{1: Fetal presentation} & Uncentered & 0.848 ($+0.28$) & 0.328 ($-0.78$) & \textbf{0.047} ($-1.00$) \\
 & Centered & 1.000 ($+0.05$) & 0.141 ($-0.89$) & 0.328 ($-0.78$) \\
\addlinespace
\multirow{2}{*}{2: Placental (sag.)} & Uncentered & 0.081 ($+0.77$) & 0.586 ($-0.39$) & 1.000 ($-0.50$) \\
 & Centered & 0.255 ($+0.67$) & 1.000 ($-0.22$) & 1.000 ($+0.22$) \\
\addlinespace
\multirow{2}{*}{3: Placental (trans.)} & Uncentered & \textbf{0.009} ($+0.95$) & 0.586 ($-0.56$) & 1.000 ($-0.06$) \\
 & Centered & 1.000 ($+0.28$) & 1.000 ($-0.50$) & 1.000 ($-0.06$) \\
\addlinespace
\multirow{2}{*}{4: Fluid measurement} & Uncentered & 0.137 ($+0.69$) & 0.547 ($-0.67$) & 1.000 ($-0.44$) \\
 & Centered & 0.261 ($+0.64$) & 1.000 ($-0.39$) & 1.000 ($0.00$) \\
\addlinespace
\multirow{2}{*}{5: Stomach} & Uncentered & 0.848 ($+0.21$) & 0.586 ($-0.56$) & 1.000 ($-0.06$) \\
 & Centered & 0.261 ($+0.62$) & 1.000 ($-0.33$) & 1.000 ($-0.11$) \\
\addlinespace
\multirow{2}{*}{6: Bladder} & Uncentered & 0.529 ($+0.46$) & 0.547 ($-0.67$) & 1.000 ($-0.28$) \\
 & Centered & 1.000 ($+0.05$) & 1.000 ($-0.33$) & 0.977 ($-0.56$) \\
\midrule
\multirow{2}{*}{Overall} & Uncentered & 0.353 ($+0.46$) & \textbf{0.023} ($-1.00$) & 0.353 ($-0.44$) \\
 & Centered & 0.467 ($+0.41$) & 0.117 ($-0.83$) & 0.467 ($-0.33$) \\
\bottomrule
\end{tabular}
\end{table}

\begin{table}[H]
\centering\small
\caption{Statistical comparisons for rotational error, comparing each mobile configuration with Magic Leap. Cells report Holm-corrected $p$ values with matched-pairs rank-biserial correlations ($r_{\mathrm{rb}}$) in brackets. R-V vs ML uses $n = 12$, while Feed and No Feed vs ML use the $n = 8$ participants who completed both sessions. Positive $r_{\mathrm{rb}}$ indicates higher values for the mobile configuration. Significant values ($p < 0.05$) are bold.}
\label{tab:S2}
\begin{tabular}{llccc}
\toprule
Task & & R-V vs ML & Feed vs ML & No Feed vs ML \\
\midrule
\multirow{2}{*}{1: Fetal presentation} & Uncentered & 0.554 ($-0.56$) & 1.000 ($+0.11$) & 1.000 ($-0.11$) \\
 & Centered & 0.904 ($-0.18$) & 1.000 ($+0.06$) & 1.000 ($+0.06$) \\
\addlinespace
\multirow{2}{*}{2: Placental (sag.)} & Uncentered & 1.000 ($+0.21$) & 0.891 ($+0.61$) & 1.000 ($+0.17$) \\
 & Centered & 0.904 ($+0.36$) & 1.000 ($+0.28$) & 1.000 ($+0.17$) \\
\addlinespace
\multirow{2}{*}{3: Placental (trans.)} & Uncentered & 1.000 ($-0.36$) & 1.000 ($-0.17$) & 1.000 ($-0.17$) \\
 & Centered & 0.815 ($-0.44$) & 1.000 ($+0.17$) & 1.000 ($-0.22$) \\
\addlinespace
\multirow{2}{*}{4: Fluid measurement} & Uncentered & 1.000 ($+0.05$) & 1.000 ($-0.33$) & 1.000 ($-0.22$) \\
 & Centered & 0.904 ($+0.31$) & 1.000 ($+0.28$) & 1.000 ($+0.39$) \\
\addlinespace
\multirow{2}{*}{5: Stomach} & Uncentered & 1.000 ($+0.33$) & 1.000 ($-0.44$) & 1.000 ($-0.39$) \\
 & Centered & 0.554 ($+0.56$) & 1.000 ($-0.06$) & 1.000 ($-0.22$) \\
\addlinespace
\multirow{2}{*}{6: Bladder} & Uncentered & 0.881 ($-0.46$) & 1.000 ($+0.11$) & 1.000 ($+0.06$) \\
 & Centered & 0.647 ($-0.51$) & 1.000 ($-0.22$) & 1.000 ($0.00$) \\
\midrule
\multirow{2}{*}{Overall} & Uncentered & 1.000 ($-0.08$) & 1.000 ($-0.11$) & 1.000 ($-0.39$) \\
 & Centered & 1.000 ($+0.10$) & 1.000 ($0.00$) & 1.000 ($-0.11$) \\
\bottomrule
\end{tabular}
\end{table}

\begin{table}[H]
\centering\small
\caption{Statistical comparisons for active task duration, comparing each mobile configuration with Magic Leap. Cells report Holm-corrected $p$ values with matched-pairs rank-biserial correlations ($r_{\mathrm{rb}}$) in brackets. R-V vs ML uses $n = 12$, while Feed and No Feed vs ML use the $n = 8$ participants who completed both sessions. Positive $r_{\mathrm{rb}}$ indicates higher values for the mobile configuration. Significant values ($p < 0.05$) are bold.}
\label{tab:S3}
\begin{tabular}{lccc}
\toprule
Task & R-V vs ML & Feed vs ML & No Feed vs ML \\
\midrule
1: Fetal presentation & 1.000 ($+0.03$) & 1.000 ($-0.44$) & 1.000 ($-0.17$) \\
2: Placental (sag.) & 1.000 ($+0.23$) & 0.328 ($-0.78$) & 1.000 ($-0.06$) \\
3: Placental (trans.) & \textbf{0.015} ($+0.92$) & 1.000 ($+0.28$) & 1.000 ($+0.50$) \\
4: Fluid measurement & \textbf{0.034} ($+0.85$) & 1.000 ($-0.28$) & 1.000 ($-0.33$) \\
5: Stomach & 0.170 ($+0.67$) & 1.000 ($-0.44$) & 1.000 ($-0.22$) \\
6: Bladder & 1.000 ($+0.28$) & 0.977 ($-0.56$) & 1.000 ($-0.44$) \\
\midrule
Overall & 0.103 ($+0.69$) & 0.297 ($-0.61$) & 0.297 ($-0.61$) \\
\bottomrule
\end{tabular}
\end{table}

\begin{table}[H]
\centering\small
\caption{Positional error, rotational error and active task duration across all four conditions, per task and overall (mean $\pm$ SD, $n = 12$ per condition). Statistical comparisons for the V-V configurations in Tables~\ref{tab:S1} to \ref{tab:S3} use the eight participants who completed both sessions, so their means differ slightly from those shown here.}
\label{tab:S4}
\resizebox{\textwidth}{!}{%
\begin{tabular}{llcccc}
\toprule
Task & & Magic Leap & Mobile R-V & Mobile V-V Feed & Mobile V-V No Feed \\
\midrule
\multicolumn{6}{l}{\textit{a) Positional error (mm)}} \\
\multirow{2}{*}{1: Fetal presentation} & Uncentered & $19.8 \pm 18.1$ & $19.2 \pm 8.3$ & $10.3 \pm 3.8$ & $10.9 \pm 5.9$ \\
 & Centered & $12.2 \pm 8.1$ & $11.7 \pm 3.6$ & $7.8 \pm 3.5$ & $7.5 \pm 2.5$ \\
\multirow{2}{*}{2: Placental (sag.)} & Uncentered & $14.4 \pm 4.6$ & $22.2 \pm 11.1$ & $11.2 \pm 3.5$ & $13.3 \pm 6.2$ \\
 & Centered & $6.6 \pm 1.7$ & $9.1 \pm 2.9$ & $5.8 \pm 1.3$ & $6.9 \pm 2.7$ \\
\multirow{2}{*}{3: Placental (trans.)} & Uncentered & $20.4 \pm 6.5$ & $34.5 \pm 9.5$ & $15.3 \pm 3.6$ & $21.4 \pm 10.2$ \\
 & Centered & $12.0 \pm 5.1$ & $15.7 \pm 9.6$ & $10.2 \pm 3.4$ & $13.2 \pm 3.7$ \\
\multirow{2}{*}{4: Fluid measurement} & Uncentered & $25.7 \pm 10.3$ & $32.5 \pm 8.3$ & $19.3 \pm 5.9$ & $22.5 \pm 10.8$ \\
 & Centered & $16.3 \pm 10.0$ & $22.7 \pm 9.4$ & $13.7 \pm 3.4$ & $15.2 \pm 2.6$ \\
\multirow{2}{*}{5: Stomach} & Uncentered & $20.9 \pm 11.3$ & $24.1 \pm 9.6$ & $17.9 \pm 10.6$ & $19.5 \pm 9.3$ \\
 & Centered & $9.9 \pm 4.7$ & $13.0 \pm 2.8$ & $9.3 \pm 3.0$ & $10.4 \pm 3.7$ \\
\multirow{2}{*}{6: Bladder} & Uncentered & $20.9 \pm 11.6$ & $28.1 \pm 10.9$ & $17.7 \pm 8.4$ & $18.0 \pm 8.9$ \\
 & Centered & $13.9 \pm 9.3$ & $14.8 \pm 4.7$ & $11.7 \pm 4.8$ & $11.4 \pm 3.0$ \\
\multirow{2}{*}{Overall} & Uncentered & $22.9 \pm 9.5$ & $28.2 \pm 8.4$ & $16.3 \pm 5.0$ & $18.2 \pm 7.8$ \\
 & Centered & $14.6 \pm 7.1$ & $16.7 \pm 4.3$ & $10.6 \pm 2.4$ & $11.4 \pm 2.0$ \\
\midrule
\multicolumn{6}{l}{\textit{b) Rotational error ($^\circ$)}} \\
\multirow{2}{*}{1: Fetal presentation} & Uncentered & $7.9 \pm 3.4$ & $6.0 \pm 1.3$ & $8.0 \pm 2.9$ & $7.3 \pm 2.4$ \\
 & Centered & $4.6 \pm 1.6$ & $4.3 \pm 1.6$ & $4.6 \pm 1.8$ & $4.7 \pm 1.8$ \\
\multirow{2}{*}{2: Placental (sag.)} & Uncentered & $6.4 \pm 3.3$ & $7.2 \pm 2.9$ & $7.3 \pm 2.9$ & $6.6 \pm 2.8$ \\
 & Centered & $3.5 \pm 1.5$ & $4.0 \pm 1.6$ & $3.2 \pm 1.4$ & $3.6 \pm 1.5$ \\
\multirow{2}{*}{3: Placental (trans.)} & Uncentered & $22.3 \pm 9.1$ & $18.1 \pm 8.2$ & $20.9 \pm 9.4$ & $21.4 \pm 6.3$ \\
 & Centered & $17.1 \pm 7.7$ & $13.6 \pm 7.8$ & $16.1 \pm 6.7$ & $16.1 \pm 4.3$ \\
\multirow{2}{*}{4: Fluid measurement} & Uncentered & $11.8 \pm 8.2$ & $10.3 \pm 5.1$ & $11.8 \pm 3.4$ & $10.5 \pm 3.2$ \\
 & Centered & $8.8 \pm 8.6$ & $8.0 \pm 4.1$ & $10.0 \pm 2.5$ & $8.2 \pm 2.2$ \\
\multirow{2}{*}{5: Stomach} & Uncentered & $9.5 \pm 6.0$ & $10.6 \pm 4.9$ & $8.2 \pm 3.2$ & $8.5 \pm 4.5$ \\
 & Centered & $5.6 \pm 3.7$ & $8.0 \pm 3.9$ & $4.8 \pm 2.4$ & $4.9 \pm 2.0$ \\
\multirow{2}{*}{6: Bladder} & Uncentered & $12.5 \pm 4.3$ & $10.6 \pm 3.7$ & $12.1 \pm 3.4$ & $10.3 \pm 3.2$ \\
 & Centered & $10.7 \pm 2.4$ & $8.7 \pm 2.8$ & $10.2 \pm 2.6$ & $9.8 \pm 3.5$ \\
\multirow{2}{*}{Overall} & Uncentered & $11.0 \pm 3.4$ & $10.6 \pm 3.0$ & $11.4 \pm 2.5$ & $10.4 \pm 2.2$ \\
 & Centered & $8.0 \pm 2.1$ & $8.1 \pm 2.2$ & $8.5 \pm 1.5$ & $7.9 \pm 1.7$ \\
\midrule
\multicolumn{6}{l}{\textit{c) Active task duration (s)}} \\
1: Fetal presentation & & $23.5 \pm 13.1$ & $24.6 \pm 9.6$ & $20.3 \pm 12.5$ & $21.4 \pm 11.7$ \\
2: Placental (sag.) & & $10.9 \pm 5.3$ & $13.8 \pm 8.3$ & $6.3 \pm 2.0$ & $7.3 \pm 3.7$ \\
3: Placental (trans.) & & $10.2 \pm 3.4$ & $21.5 \pm 13.4$ & $13.1 \pm 7.8$ & $14.1 \pm 6.8$ \\
4: Fluid measurement & & $27.2 \pm 6.4$ & $69.7 \pm 84.4$ & $22.1 \pm 8.9$ & $21.5 \pm 8.9$ \\
5: Stomach & & $24.3 \pm 12.8$ & $40.6 \pm 22.8$ & $16.3 \pm 7.3$ & $21.8 \pm 15.6$ \\
6: Bladder & & $49.4 \pm 54.4$ & $58.6 \pm 42.3$ & $39.9 \pm 33.7$ & $34.8 \pm 34.8$ \\
Overall & & $145.5 \pm 67.5$ & $228.7 \pm 122.6$ & $117.9 \pm 50.7$ & $120.9 \pm 55.6$ \\
\bottomrule
\end{tabular}%
}
\end{table}

\end{document}